\documentclass[doublespacing]{elsart}

\usepackage{icarus}

\usepackage{natbib}
\usepackage{longtable}

\bibpunct{(}{)}{;}{a}{,}{,}

\usepackage{graphicx}

\usepackage{amssymb}

\newcommand{\BibTeX}{ \textrm{B\kern-.05em\textsc{i\kern-.025em b}\kern-.08em
    T\kern-.1667em\lower.7ex\hbox{E}\kern-.125emX} }

\begin{document}

\begin{frontmatter}



\title{The Distribution of Basaltic Asteroids in the Main Belt}


\author[ifa]{Nicholas A. Moskovitz}, 
\author[ifa]{Robert Jedicke},
\author[eric,ab]{Eric Gaidos},
\author[ifa]{Mark Willman},
\author[dave]{David Nesvorn\'{y}},
\author[ron]{Ronald Fevig},
\author[zel]{\v{Z}eljko Ivezi\'{c}}

\address[ifa]{Institute for Astronomy, 2680 Woodlawn Drive, Honolulu, HI 96822 (U.S.A)}
\address[eric]{Department of Geology and Geophysics, University of Hawaii, POST 701, 1680 East-West Road, Honolulu, HI 96822 (U.S.A)}
\address[ab]{NASA Astrobiology Institute, University of Hawaii, PSB 213, 2565 McCarthy Mall, Honolulu, HI 96822 (U.S.A)}
\address[dave]{Department of Space Studies, Southwest Research Institute,1050 Walnut Street, Suite 400, Boulder, CO 80302 (U.S.A.)}
\address[ron]{Lunar \& Planetary Laboratory, University of Arizona,1629 University Blvd., Tucson, AZ 85721-0092 (U.S.A.)}
\address[zel]{Department of Astronomy, University of Washington, PO Box 351580, Seattle, WA 98195-1580 (U.S.A.)}

\begin{center}
\scriptsize
Copyright \copyright\ 2007 Nicholas A. Moskovitz
\end{center}


%
%
%
%
%


\end{frontmatter}



\begin{flushleft}
\vspace{1cm}
Number of pages: \pageref{lastpage} \\
Number of tables: \ref{lasttable}\\
Number of figures: \ref{lastfig}\\
\end{flushleft}


\begin{pagetwo}{Basaltic Asteroids in the Main Belt}

Nicholas A. Moskovitz\\
Institute for Astronomy\\
2680 Woodlawn Drive\\
Honolulu, HI 96822, USA. \\
\\
Email: nmosko@ifa.hawaii.edu\\
Phone: (808) 956-6700 \\

\end{pagetwo}

\begin{abstract}

We present the observational results of a survey designed to target and detect asteroids whose colors are similar to those of Vesta family members and thus may be considered as candidates for having a basaltic composition. Fifty basaltic candidates were selected with orbital elements that lie outside of the Vesta dynamical family. Optical and near-infrared spectra were used to assign a taxonomic type to 11 of the 50 candidates. Ten of these were spectroscopically confirmed as V-type asteroids, suggesting that most of the candidates are basaltic and can be used to constrain the distribution of basaltic material in the Main Belt.

Using our catalog of V-type candidates and the success rate of the survey, we calculate unbiased size-frequency and semi-major axis distributions of V-type asteroids. These distributions, in addition to an estimate for the total mass of basaltic material, suggest that Vesta was the predominant contributor to the basaltic asteroid inventory of the Main Belt, however scattered planetesimals from the inner Solar System ($a<2.0$ AU) and other partially/fully differentiated bodies likely contributed to this inventory. In particular, we infer the presence of basaltic fragments in the vicinity of asteroid 15 Eunomia, which may be derived from a differentiated parent body in the middle Main Belt ($2.5<a<2.8$). We find no asteroidal evidence for a large number of previously undiscovered basaltic asteroids, which agrees with previous theories suggesting that basaltic fragments from the $\sim100$ differentiated parent bodies represented in meteorite collections have been "battered to bits" [Burbine, T.H., Meibom, A., Binzel, R.P., 1996. Mantle material in the Main Belt: Battered to bits? Met. \& Planet. Sci. 31, 607].

\end{abstract}



\section{Introduction \label{sec.intro}}

The surface properties of asteroids as determined by spectroscopic observations are clues to the environmental conditions present in the early Solar System. Asteroids with basaltic material on their surface are interpreted as tracers of bodies whose interiors reached the melting temperature of silicate rock and subsequently differentiated \citep{Gaffey93b,Gaffey02}. Until recently Vesta and the dynamically associated Vestoids were the only asteroids with observed surface basalt \citep{McCord70,Binzel93}. A small number of basaltic asteroids have now been found that may derive from other differentiated parent bodies \citep{Lazzaro00,Hammergren06,Binzel07,Roig08}. Other asteroids and asteroid families may also represent differentiated parent bodies [e.g. Flora \citep{Gaffey84}, Eunomia \citep{Reed97,Nathues05}, Eos \citep{Mothe05}, Hebe \citep{Gaffey98}, and Thetis, Agnia and Merxia \citep{Sunshine04}], however it is not clear why so little basaltic material from these objects remains while the Vestoids are one of the most prominent families in the Main Belt. Two possibilities are that the Vestoids were produced relatively recently and that non-Vestoid basaltic fragments were heavily comminuted to sizes below our current observational limits \citep{Davis89,Burbine96,Davis02,Bottke05b} or that some space weathering process has erased the diagnostic spectral signatures of surface basalts \citep[e.g.][]{Wetherill88}.

In addition to asteroidal evidence for the widespread occurrence of differentiation, oxygen isotope analyses of the iron meteorites suggests that as many as $\sim60$ differentiated parent bodies once existed \citep{Chabot06,Burbine02}, some of which were more than 500 km in size \citep{Yang07}. Furthermore, the HED basaltic achondrites are not isotopically homogenous and may represent more than one differentiated parent body \citep{Yamaguchi01}. 

These populations suggest that basaltic material was common throughout the inner Solar System, however it is unclear whether there currently exists any basaltic vestige of the numerous differentiated parent bodies that once existed. To provide insight on these issues we have determined an unbiased distribution of V-type asteroids in the Main Belt. Other studies have calculated unbiased distributions of asteroid types \citep[e.g.][]{Gradie82,Gradie89,Bus02}, but they have been restricted to the largest and brightest asteroids and did not consider the presence of V-type asteroids beyond 2.5 AU. Our survey provides constraints on the number of basaltic asteroids up to an absolute magnitude of 17.2, which is an increase of roughly 3 orders of absolute magnitude compared to these previous studies and therefore probes for basaltic fragments that may represent the upper size limit to a heavily comminuted population.

Throughout this work we use specific terms related to the taxonomic classification, surface mineralogy and dynamical properties of asteroids. V-type is a taxonomic distinction based on spectral criteria defined by \citet{Bus02}. Any object with photometric colors consistent with a V-type designation is a V-type candidate. Basaltic asteroids are differentiated bodies or fragments of differentiated bodies with basalt on at least part of their surface. Basaltic asteroids are characterized based on interpreted mineralogy from spectroscopic data. An object with photometric colors indicative of a basaltic composition is a basaltic candidate. For this work we consider a V-type taxonomy to be directly attributable to a basaltic composition, therefore the terms V-type and basaltic are interchangeable. Vestoids are a dynamically-defined sub-population of V-types thought to have originated from the Vesta parent body at least 1 billion years ago \citep{Marzari96}. Both spectroscopic data and dynamical studies are required to classify an object as a Vestoid. Vestoids are typically interpreted as having a basaltic surface composition, however some asteroids with basaltic material on their surface are not Vestoids, e.g. Magnya \citep{Lazzaro00}. An asteroid that is dynamically associated with Vesta but has not been spectroscopically observed is called a dynamical Vestoid. 

In the early Solar System the likely heat source for metal-silicate differentiation and the production of basalt was the decay of short-lived radioactive isotopes such as $^{26}$Al and $^{60}$Fe \citep{Grimm93,Goswami00,Tachibana03}. The half-lives of these isotopes are 0.73 and 1.5 Myr respectively, thus under this scenario a differentiated planetesimal or embryo must have accreted within a few half-lives ($< 3$ Myr after the onset of decay) to reach silicate melting temperatures ($>1000$ K). An alternative heat source for material in the early Solar System may have been electromagnetic inductive heating by an intense solar wind \citep{Herbert91,Gaffey02}, however this mechanism is poorly constrained \citep{McSween02} and may be incompatible with observations of T Tauri winds \citep{Wood91}.

The increase of accretion time-scale with heliocentric distance implies that planetesimals from the terrestrial planet region of the Solar System incorporated a high abundance of live radionuclides and may have differentiated, whereas planetesimals that formed in the vicinity of the gas giants accreted too slowly to have melted. It is uncertain where the boundary between these two regions lay. \citet{Bottke06} suggest that differentiation was primarily confined to the inner Solar System ($a<2.0$ AU) so that the iron meteorites that reach the Earth today trace differentiated planetesimals that were scattered into the Main Belt from smaller heliocentric distances. It is unclear whether the presence of non-Vestoid basaltic asteroids and partially melted asteroids in the outer belt [e.g.~Eos and Eunomia, \citet{Mothe05,Reed97,Nathues05}] provide evidence for the in-situ differentiation of material in the Main Belt or can be explained by the \citet{Bottke06} scenario.

Basaltic melts derived from a chondritic precursor will contain pyroxenes, olivine and plagioclase. Pyroxenes are spectroscopically characterized by deep and broad 1- and 2-$\mu m$ absorption features. Figure \ref{fig.basalt} shows optical and near-infrared spectra of basaltic asteroids 38070 (1999 GG2) and 33881 (2000 JK66), which are two of our basaltic candidates (\S\ref{sec.selection}). The visible spectrum was obtained with ESI \citep{Sheinis02} on Keck II and the near-infrared with SpeX \citep{Rayner03} on IRTF. Also plotted is a RELAB (Reflectance Experiment Laboratory) spectrum of a basaltic eucrite meteorite. These objects clearly show the absorption features at 1 and 2 $\mu m$. The depth, center and shape of these features are dictated by temperature, grain size and calcium content \citep{Hardersen04} and are attributable to Fe$^{2+}$ ionic transitions within the pyroxene crystal structure \citep{Sunshine04}. Olivine and plagioclase also show Fe$^{2+}$ absorption features around 1 $\mu m$. 

Although the spectral features of basaltic material dominate in the near-infrared, optical data ($0.4-0.95~\mu m$) are easier to obtain for faint asteroids and can be diagnostic of such surface material. Thus, we primarily relied on optical taxonomic classification as a way of selecting and characterizing our target asteroids. 

Recent observational and theoretical work \citep{Lazzaro00,Hammergren06,Bottke06,Binzel07, Nesvorny08,Roig08} along with the data presented here suggests that there exists non-Vestoid basaltic asteroids throughout the Main Belt that have yet to be discovered. We have placed new constraints on the overall population of these rare objects by investigating relatively faint asteroids that are not typically included in spectroscopic surveys. One of the goals of all research related to the extent of thermal alteration in the early Solar System is to understand how many differentiated parent bodies are represented by the Main Belt asteroids. This work is a small step towards that understanding.

The remaining sections of this paper are organized as follows: in \S\ref{sec.selection} we outline the procedure of selecting target asteroids for our observational campaign. In \S\ref{sec.obs} and \S\ref{sec.reduct} we discuss our observational procedure and data reduction techniques. In \S\ref{sec.dist} we use our observational results to derive a size-frequency distribution of basaltic asteroids in the Main Belt. We use this result to constrain the semi-major axis distribution and the total mass of basaltic asteroids. In \S\ref{sec.disc} we discuss and summarize these results.

\section{Selecting Basaltic Asteroid Candidates \label{sec.selection}}

\begin{figure}[p!]
\begin{center}
\includegraphics[width=8.5cm]{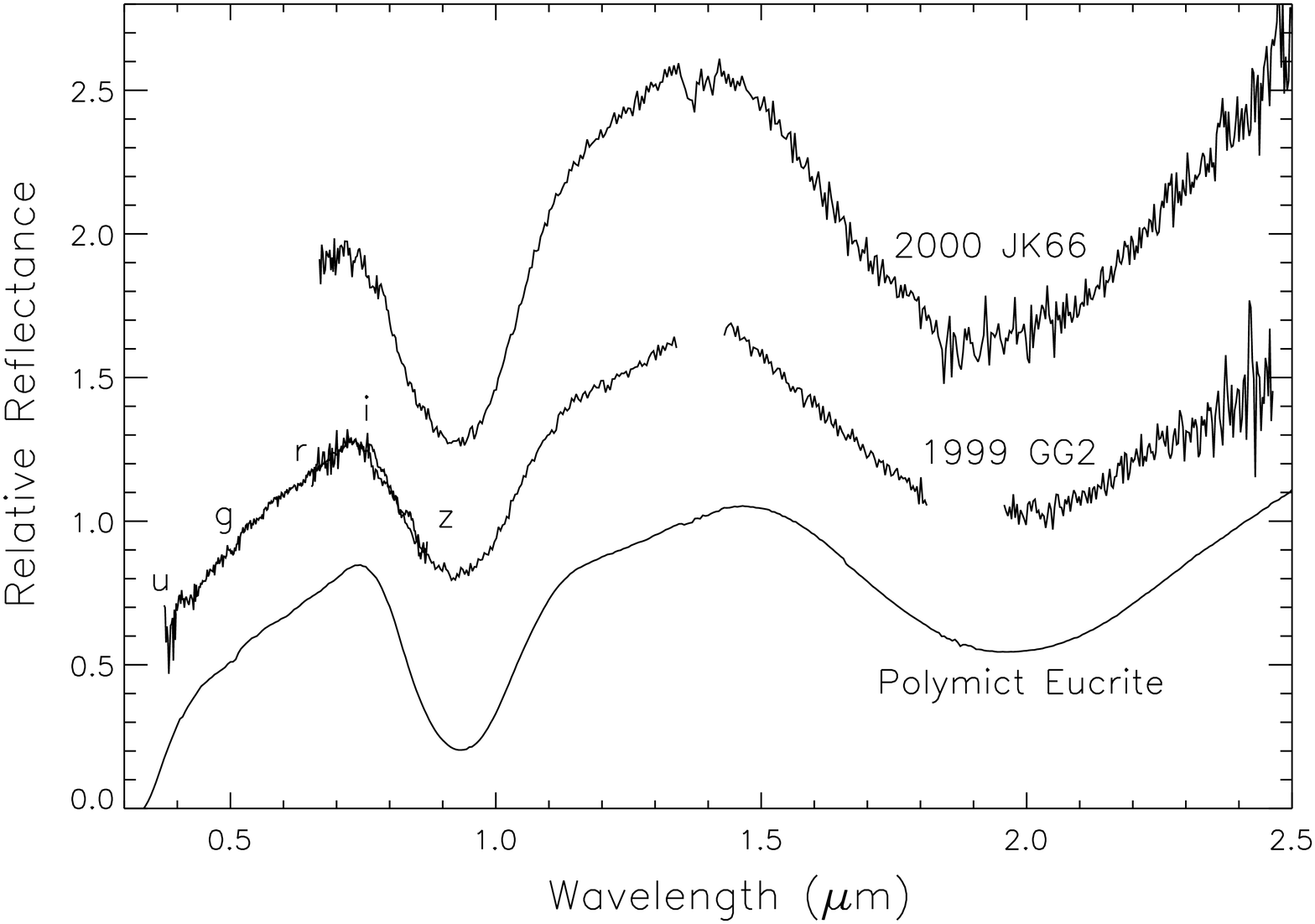}
\caption[]{
Optical (ESI on Keck II) and near-infrared (SpeX on NASA's Infrared Telescope Facility) spectra of basaltic asteroids 38070 (1999 GG2) and 33881 (2000 JK66) and a laboratory spectrum of polymict eucrite Y74450 (acquired at the NASA RELAB facility, Brown University by Takahiro Hiroi). These data clearly show the broad 1 and 2 $\mu m$ absorption features that are characteristic of basaltic material. The band centers of the SDSS {\it ugriz} filters are indicated on the spectrum of 38070. The spectra have been offset by increments of $\pm0.5$ units relative to the 1999 GG2 spectrum.}
\label{fig.basalt}
\label{lastfig}
\end{center}
\end{figure}

We selected candidate basaltic asteroids from the photometric data in the third release of the Sloan Digital Sky Survey \citep[SDSS,][]{York00} Moving Object Catalog \citep[MOC,][]{Ivezic02}. Our targets were chosen based upon their orbital elements and a statistically significant similarity to the {\it ugriz} photometric colors of the Vestoid population. The {\it ugriz} bands span a wavelength range from approximately $0.35 - 0.90~\mu m$. These five filters provide a crude optical spectrum that can indicate the presence of basaltic spectral features (Fig. \ref{fig.basalt}).

The 204,305 moving objects in the MOC were filtered down to $\sim$40,000 by removing multiple observations of single bodies, objects not matched to known asteroids in the ASTORB orbital element database \citep{Juric02,Bowell07}, and objects with any photometric error larger than one magnitude. If  this constraint on the photometric errors was more rigid, then the number of final candidates would have decreased. This would have reduced the statistical significance of our result and made it more difficult to coordinate observing runs with observable targets.

The SDSS magnitudes published in the MOC are typically denoted by {\it u'g'r'i'z'}. A transform (from the SDSS website, http://www.sdss.org/dr6/) is applied to these magnitudes so that solar correction in the AB system can be performed:
\begin{eqnarray}
u = u' - 0.04 \nonumber \\
g = g' + 0.060~(g' - r'-0.53)\nonumber \\
r = r' + 0.035~(r' - i'-0.21) \nonumber \\
i=i' + 0.041~(r' - i'-0.21) \nonumber \\
z = z' -0.030~(i' - z' -0.09) +0.02
\label{eqn.ABtrans}
\end{eqnarray}
Four color combinations were calculated ($u-g,~g-r,~r-i,~i-z$) and solar corrected for each object in our filtered version of the MOC. The adopted solar colors were $u-g=1.43$, $g-r=0.44$, $r-i=0.11$ and $i-z=0.03$ and were subtracted from each color permutation. In addition, two principal component colors were calculated for each asteroid \citep{Nes05}:
\begin{eqnarray}
PC1 = 0.396~(u-g) + 0.553~(g-r) + 0.567~(g-i) + 0.465~(g-z) \nonumber \\
PC2 = -0.819~(u-g) + 0.017~(g-r) + 0.090~(g-i) + 0.567~(g-z)
\label{eqn.PCs}
\end{eqnarray}
These linear combinations define a set of axes that maximize the separation between S- and C-type asteroids \citep{Nes05}.

\begin{figure}[p!]
\begin{center}
\includegraphics[width=8.5cm]{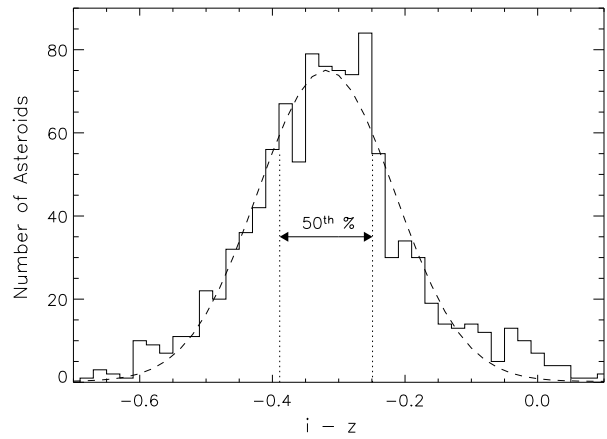}
\caption[]{
$i-z$ color histogram of the 1051 dynamical Vestoids that are included in the SDSS MOC. A gaussian fit to the distribution is shown by the dashed curve. The range of colors that correspond to the $50^{th}$ percentile of dynamical Vestoids (defined symmetrically relative to the center of the gaussian) is indicated.}
\label{fig.vest_hist}
\label{lastfig}
\end{center}
\end{figure}

Using a Hierarchical Clustering Method \citep[HCM,][]{Zappala90} and a velocity cutoff of 70 m/s we identified 5575 dynamical Vestoids from a 2005 version of the AstDys database (http://hamilton.dm.unipi.it/cgi-bin/astdys/astibo). This cutoff does not include the 365 m/s escape velocity \citep{Binzel93} from Vesta's potential well. Of these dynamical Vestoids, 1051 are included in the MOC. Certainly some fraction of these are non-V-type interlopers, however Fig. \ref{fig.vest_hist} shows that the colors of these 1051 objects are distributed with a low dispersion which suggests that the number of interlopers is small and that this population can be used to represent the photometric colors of V-type asteroids. 

We fit histograms of these colors with gaussian profiles (Fig. \ref{fig.vest_hist}) to define color ranges corresponding to percentile groups within the dynamical Vestoid population. These ranges were defined symmetrically relative to the fitted centroid of the profiles. For instance, the $50^{th}$ percentile of dynamical Vestoids in the MOC have $u-g$ colors between  0.17 and 0.35 with a most probable value of 0.26. We used the $50^{th}$ percentile for each of the 6 colors to statistically define the SDSS photometric properties of candidate basaltic asteroids. Table \ref{tab.percentile} shows the adopted range for each color. Basaltic candidates were selected as objects whose colors overlap all 6 of these ranges.

\begin{figure}[p!]
\begin{center}
\includegraphics[width=8.5cm]{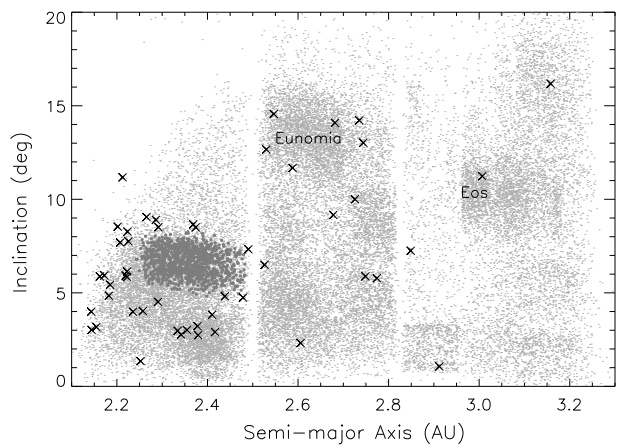}
\caption[]{
Osculating orbital elements of Vestoids (grey circles), MOC objects (small dots) and basaltic candidates ($\times$). The Eunomia and Eos families are indicated. The apparent overlap of some candidates with the Vestoids is due to projection along the unplotted eccentricity axis. Note that there are at least 6 candidates in the vicinity of the Eunomia family.
}
\label{fig.orbits}
\label{lastfig}
\end{center}
\end{figure}

Dynamical constraints were then applied to objects that met the photometric criteria. Asteroids were excluded from the candidate list if their orbital properties overlapped the region of phase space occupied by the dynamical Vestoids. This dynamical constraint was applied to the osculating values of semi-major axis, eccentricity and inclination. Proper orbital elements would have been preferred but these are available for only a subset of MOC objects. The extent of phase space occupied by the 1051 dynamical Vestoids in the MOC is 2.22 - 2.48 AU in semi-major axis, 0.029 - 0.169 in eccentricities, and 4.82 - 8.26$^\circ$ in inclination. These photometric and dynamical constraints produced 50 targets for spectroscopic follow-up (Table \ref{tab.candidates}, Fig. \ref{fig.orbits}).

Table \ref{tab.candidates} indicates the relative ejection velocity, $\Delta v$, required for each of the candidates to have originated from the surface of Vesta. Simplified forms of Gauss's equations can be used to express an asteroid's change of velocity in three components (tangential, radial and perpendicular to the orbital plane) relative to another body \citep[Equations 4, 5 and 6 in][]{Zappala96}. The net relative ejection velocity, $\Delta v$, is calculated by adopting these authors' values for the true anomaly and argument of perihelion of the Vesta family at the time of its collisional formation (see their Fig. 12) and then adding the three components in quadrature. The 365 m/s escape velocity from Vesta \citep{Binzel93} is not taken into account for these calculations. With the exception of two objects [28517 (2000 DD7) and 156914 (2003 FB31)] all of the ejection velocities are greater than the $\sim0.6$ km/s that is expected for multi-kilometer Vestoid fragments \citep{Asphaug97}. 

It should be noted that calculating $\Delta v$ for a pair of asteroids can be a useful metric for quantifying dynamical proximity, however dynamical effects such as Yarkovsky drift and resonance scattering can produce values of $\Delta v$ for members of the same family that are much larger than the sub-km/s velocities expected from HCM family definitions \citep[e.g.][]{Zappala95,Zappala96} and from SPH simulations of family forming impacts \citep[e.g.][]{Asphaug97,Nesvorny06}. For instance, \citet{Nesvorny08} show that Vestoids with $\Delta v>2$ km/s are likely and \citet{Roig08} suggest that asteroid 40521 (1999 RL95) is a fragment from Vesta's surface even though its $\Delta v=2.3$ km/s. Part of this problem is rooted in the HCM technique: only objects that ``stand out" from the background population of asteroids can be identified as members of a collisional family. In reality the dynamical boundaries of collisional families are not so clearly defined and certainly extend into the background population and even other families. 

For these reasons it is likely that our HCM-defined dynamical boundaries do not encompass the full extent of the Vestoid family, a presumption that is supported by the presence of basaltic candidates near these boundaries (Fig. \ref{fig.orbits}). However, the inclusion of these objects as candidates does not dramatically affect our analysis or conclusions. Using a larger cutoff velocity ($>70$ m/s) to expand the dynamical boundaries would connect all of the asteroids in the inner Main Belt ($a<2.5$ AU) and would therefore neglect the possibility of selecting basaltic asteroids from other Vesta-like differentiated parent bodies. The use of a ``small" cutoff velocity (i.e. one that does not encompass the entire Vestoid population) is conservative in the sense that it makes the amount of supposedly non-Vestoid basaltic asteroids larger and thus contributes to raising the upper limit for the total mass of basaltic asteroids (\S\ref{sec.dist}.2).

Other studies have employed similar color-selection techniques for identifying V-type candidates in the MOC \citep{Roig06,Hammergren06,Binzel07}. \citet{Roig06} used two principal components and $i-z$ colors to place photometric constraints on the MOC, generating over 250 non-Vestoid basaltic candidates. Sixteen of our 50 candidates are included in the \citet{Roig06} list (Table \ref{tab.candidates}). The  lack of overlap is due to the different selection techniques. \citet{Roig06} used only three colors to select candidates, but applied more stringent criteria on the acceptable photometric error bars. Our technique used 6 colors but applied more relaxed constraints on the photometric errors. \citet{Hammergren06} selected V-type candidates based on $i-z$ colors which are diagnostic of the 1 $\mu$m basaltic absorption feature (Fig. \ref{fig.basalt}) and \citet{Binzel07} developed a system to translate the photometric measurements of MOC objects into taxonomic classifications based on additional spectroscopic data \citep{Bus02}. The last three columns in Table \ref{tab.candidates} indicate whether our candidates were also identified as V-type candidates by these other groups.

It is unclear whether one of these methods is more effective at taxonomic prediction than the others. Our method and the \citet{Roig06} technique produce similar results: 11 of 12 \citet{Roig06} candidates and 10 out of 11 of our targets are spectroscopically confirmed V-types (this work, \citet{Roig08,Roig06,Lazzaro04,Bus02}). The significant overlap between our candidates and those selected by \citet{Hammergren06} suggest that these two methods are roughly equivalent. \citet{Binzel07} and Masi et al. (2008, submitted to A\&A) show that their technique is $\sim80\%$ effective at predicting taxonomic type.

\section{Spectroscopic Observations \label{sec.obs}}

We have obtained spectra of 9 basaltic candidates.\footnote[1]{We have also observed a number of asteroids that fall just outside of the $50^{th}$ percentile color ranges (not listed in Table \ref{tab.candidates}) and asteroids that are part of the \citet{Roig06} candidate list. The taxonomic results of these observations can be found at www.ifa.hawaii.edu/$\sim$nmosko/Home/astobs.html} Table \ref{tab.obs} contains a summary of the observational circumstances for each target asteroid. In addition, the candidates 40521 (1999 RL95) and 2823 (van der Laan) have been observed as part of complementary projects performed by other groups \citep{Binzel01,Roig08}.

We used the Supernova Integral Field Spectrograph (SNIFS) \citep{Lantz04} on the University of Hawaii 2.2 meter telescope to obtain optical spectra ($0.4-95~\mu m$) of asteroids brighter than $V\sim 18.5$ magnitude. The SNIFS 6" $\times$ 6" field of view is divided into  0.4" $\times$ 0.4" spaxels (spectral pixels), which are each generated by one lens in a 225 element multi-lenslet array. The light from each spaxel is fiber-fed to a dichroic mirror that splits the light into red and blue channels. The two channels pass through separate grisms and are imaged on two 2k $\times$ 4k blue- and red-optimized CCDs. The approximate spectral resolution ($R$) of this instrument is 1000 at $0.43~\mu m$ and 1300 at $0.76~\mu m$. Bias frames, dome flats and internal arc lamp spectra are automatically obtained as part of the standard SNIFS observing routine.

The Echellette Spectrograph and Imager (ESI) on Keck II \citep{Sheinis02} was used to obtain optical spectra ($0.38-0.90~\mu m$) of objects that were too faint to observe with SNIFS. ESI was operated in the low resolution ($R\sim1000$) prism mode and with a wide slit of 6". We found that this width resulted in greater consistency between multiple spectra of the same asteroid.  To minimize the effect of atmospheric dispersion the orientation of the slit was set close to the parallactic angle at the midpoint of the observation. Bias frames and dome flat field images were taken at the beginning and end of each night. Wavelength calibration was achieved by taking spectra of internal Hg, Cu, and Xe arc lamps at the beginning and end of each night. The dispersion for ESI was found to be very stable so that one set of arc line images could have been used to calibrate all of our data. 

Near-infrared (NIR) observations were performed with SpeX \citep{Rayner03} on NASA's Infrared Telescope Facility (IRTF). The telescope was operated in a standard ABBA nod pattern and SpeX was configured in its low resolution (R=250) prism mode with a 0.8" slit for wavelength coverage from 0.8 - 2.5 $\mu$m. Internal flat fields and arc line spectra were obtained immediately after each target asteroid. The IDL-based SpeXtool package \citep{Cushing04} was used for data processing.

All targets were observed at an air mass of less than 1.3 and as near as possible to their meridian crossings. Bright field stars were used to guide the telescopes at non-sidereal rates so that the target asteroid remained centered on the slit for ESI and on the lenslet array for SNIFS. For the optical instruments we employed a maximum exposure per image of 900 seconds.  For SpeX, exposure times were limited to 120 seconds. The number of exposures was selected in an attempt to obtain a consistent $S/N\sim100$ for all asteroids.

A solar analog star was observed to perform calibrations for each asteroid spectra (i.e. solar correction, compensating for instrument response and removal of atmospheric absorption features, see \S\ref{sec.reduct}). Analogs were observed closely in time, altitude and azimuth to the asteroid. Each one-dimensional asteroid spectrum was divided by that of an analog during the reduction process. For SNIFS and SpeX data a single solar analog was selected for each asteroid. We found that this prescription produced the most consistent results, which may be related to the use of brighter rather than well established solar analogs. For ESI we followed the procedure of \citet{Bus02} and divided the asteroid spectra by the average of all analogs for a given night. The specific analog that was used for the calibration of each asteroid is given in Table \ref{tab.obs}.

\section{Data Reduction \label{sec.reduct}}

Although the reduction of asteroid spectral data varies from one instrument to the next, a number of key steps are common to all optical reductions. We outline non-standard steps here because we are not aware of an earlier comprehensive description of these processes. For further details on spectral processing of our optical data see \citep{Willman08}.

Raw two-dimensional spectral images have a number of superimposed components that must be removed to isolate one-dimensional asteroid reflectance spectra. Pixel-to-pixel variations on a CCD must be corrected by dividing data images with a normalized flat field image of a uniformly illuminated source. The normalization of this flat field must take into account the sensitivity variations on the CCD as function of position {\it and} as a function of wavelength. Some CCDs display optical fringing at red wavelengths (the ESI CCD shows fringing effects at the few percent level around 0.9 $\mu$m), which can be corrected by dividing the data images with a master fringe frame. The fringe pattern that affects data images is produced by parallel light rays originating at infinity, thus it can not be reproduced with dome flats. A normalized exposure of the twilight sky can be used to reproduce fringe patterns. The extraction of the 1D spectrum from the 2D image is performed by defining and subtracting an average background from the asteroid data. This effectively removes any atmospheric emission features. To correct for atmospheric absorption the 1D asteroid spectrum is divided by a solar analog spectrum. Division by the analog also removes the superimposed solar contribution in the measured reflectance.

For SNIFS, one-dimensional spectra from the red and blue channels are produced within a few minutes of data acquisition. These spectra are bias level-calibrated, flat field-corrected, cosmic ray-cleaned and have an assigned wavelength solution. We have developed a series of custom IDL scripts to perform the remainder of the data reduction. These scripts normalize the spectra at 0.55 $\mu$m, combine multiple exposures of a single object, join the red and blue channel data, divide the asteroid by a solar analog and finally re-bin the data to increase S/N. Around 0.5 $\mu$m, the sensitivity of the SNIFS red and blue channels drops, thus causing significant degradation of signal-to-noise. For cosmetic purposes the data in this region was removed.

Spectral images from ESI have a rotated dispersion axis relative to the x-axis of the CCD and significant spectral line curvature along the spatial direction. We corrected for these two effects by rotating the 2D images by $6.4^\circ$ clock-wise and then straightening the curvature by applying a transform map produced from arc line images. The {\it identify, reidentify, fitcoords} and {\it transform} IRAF \citep{Tody86} routines were used to perform this straightening. Once all of the spectral images were straightened the reduction proceeded following standard protocols. Dispersion solutions were determined based upon 28 different arc lines distributed throughout the wavelength range of the instrument. After dividing the asteroid spectrum by the solar analog we further divided it by a scaled model of the atmospheric absorption features that are present beyond $0.85~\mu m$ \citep{Willman08} This procedure is a fine tuning to the final asteroid spectrum and effectively eliminates any residual absorption.

\begin{figure}[p!]
\begin{center}
\includegraphics[width=17cm]{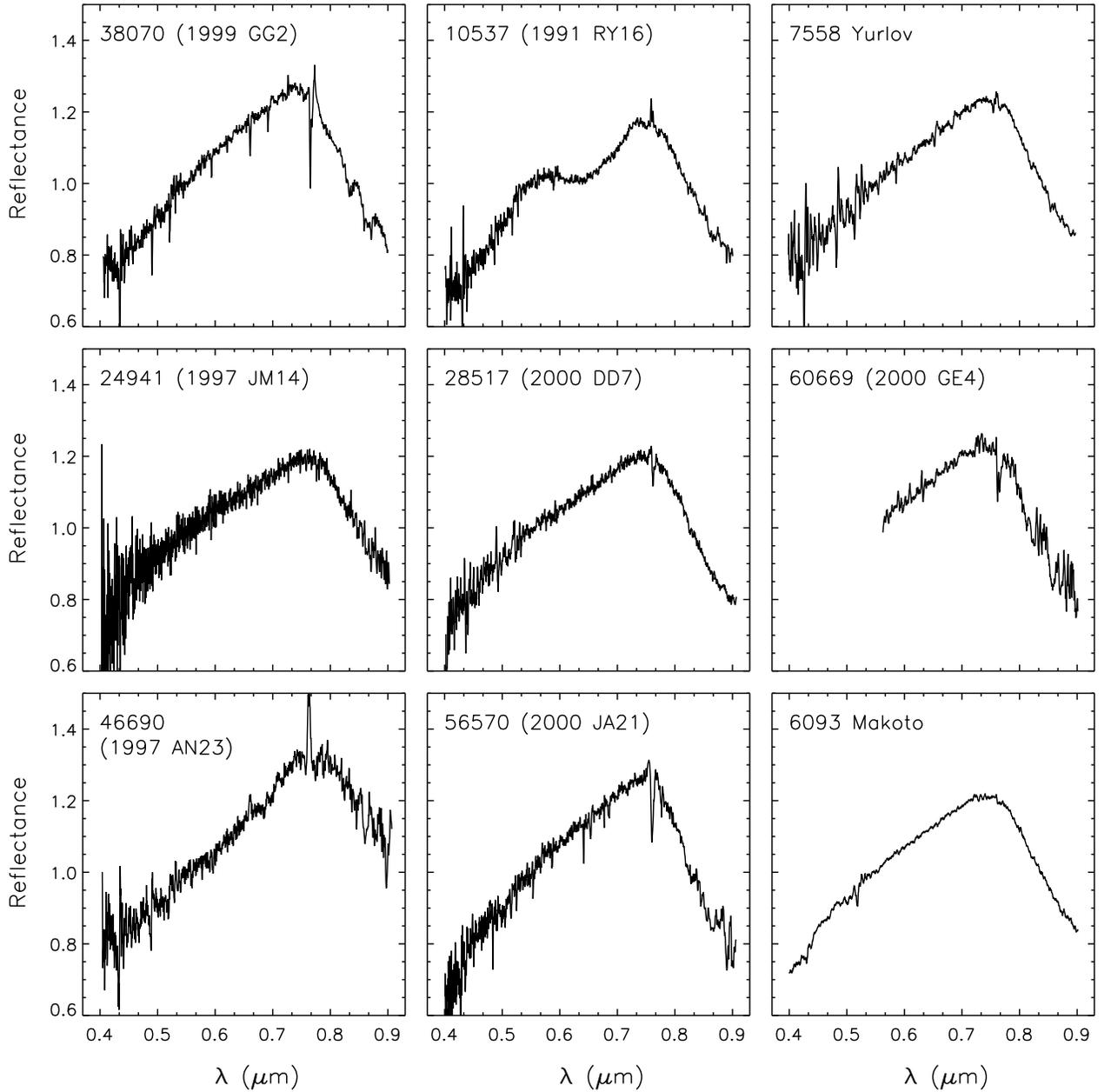}
\caption[]{Optical spectra of observed basaltic candidates. Each spectrum has been normalized at 0.55 $\mu$m. Due to instrumental problems, only the red channel data from SNIFS is shown for asteroid 60669 (2000 GE4). A spectrum of Vestoid 6093 Makoto is included for reference (obtained with SNIFS on 26 October 2006). The shallow depth of the 1 $\mu$m band in the spectrum of 1997 AN23 suggests that it is an S-type asteroid, whereas all of the others are V-types.
}
\label{fig.opt_spec}
\label{lastfig}
\end{center}
\end{figure}

Reduced optical spectra of the observed candidates are presented in Fig. \ref{fig.opt_spec}. On the night of 18 June 2007, the blue channel spectrograph on SNIFS was inoperable, thus only the red channel data are presented for asteroid 60669 (2000 GE4). A spectrum of Vestoid 6093 Makoto is included in Fig. \ref{fig.opt_spec} for reference; this asteroid is not one of our basaltic candidates. Only NIR observations were obtained for candidate 33881 (2000 JK66). The SpeX spectrum of this target is shown in Fig. \ref{fig.basalt} and clearly shows the deep 1- and 2-micron absorption bands that are indicative of a basaltic surface composition.

\begin{figure}[p!]
\begin{center}
\includegraphics[width=8.5cm]{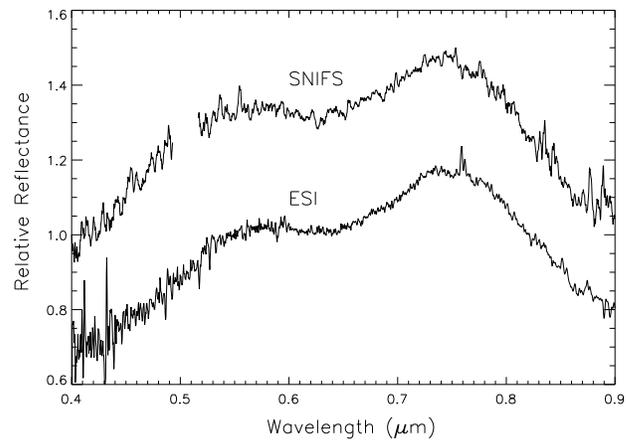}
\caption[]{ESI and SNIFS spectra of basaltic candidate asteroid 10537 (1991 RY16). The gap in the SNIFS data corresponds to the transition region between the red and blue channels. The SNIFS data have been offset for clarity.
}
\label{fig.RY16}
\label{lastfig}
\end{center}
\end{figure}

Asteroid 10537 (1991 RY16) was observed with both SNIFS and ESI (Fig. \ref{fig.RY16}) and thus allows a comparison between spectra from the two instruments. ESI appears to have better red sensitivity, however we have not tried to fine tune the SNIFS data with an atmospheric absorption model.  The slight discrepancy between these two spectra at the red-most wavelengths is likely due to imperfect correction of atmospheric features and the difference in solar analog selection. Other than these subtle differences there is very good consistency between the SNIFS and ESI data.

A number of criteria were used to discriminate a V-type asteroid from the optical data \citep{Bus02}. These included: (1) a moderate $0.4-0.7~\mu m$ slope that can include a shoulder at $\sim0.6~\mu m$ \citep{Vilas00}, (2) a peak reflectance of $\sim1.2$ at $0.75~\mu m$ when normalized at the canonical 0.55 $\mu m$ and (3) the blue edge of the $1~\mu m$ absorption feature reaching a depth of 0.8 or greater at 0.9 $\mu m$.

Asteroid types A, Q and R have similar spectral features to V-types and could become false positives in the taxonomic classification of our targets. However, errors produced by incorrectly classifying an A, Q or R-type asteroid have little bearing on our conclusions since it is likely that A- and R-types are also derived from differentiated parent bodies \citep{Marchi04,Lucey98}. Furthermore, since there are no Main Belt Q-types amongst the more than 2000 asteroids with assigned taxonomies it is improbable that any of our fifty Main Belt candidates are Q-type. 

In some cases a taxonomic distinction can be made \citep{Bus02}. A-types tend to have steeper slopes blue-ward of $0.75~\mu m$ and shallower $1.0~\mu m$ absorption features. At optical wavelengths Q-types are typically distinguished from the V-class by a smaller peak reflectance ($\sim 1.1$) and a shallower $1.0~\mu m$ absorption feature. R-type asteroids have very similar spectral features to those of the V-class, however these asteroids are exceedingly rare. It would be unlikely that we discover one of these as part of our observational program. A distinction between R- and V-types can be made based on a shallower $1~\mu m$ absorption feature for R-types. 

S-complex asteroids are the most common contaminant in our list of candidates. This is attributable to a few factors. By number, they are the second most abundant asteroid type in the Main Belt. In addition, the difference between S and V-class asteroids can be obscured by the relatively low signal-to-noise of the MOC measurements and the statistical method by which we have selected these targets. Fortunately S-types are easy to spectroscopically distinguish from V-types. For an S-type, the typical slope blue-ward of $0.75~\mu m$ is less steep and the $1~\mu m$ absorption feature is shallower ($\sim0.9$). 

The spectral classifications of our observed basaltic candidates are listed in Table \ref{tab.obs}. The truncated spectrum of 60669 (2000 GE4) was sufficient to determine its taxonomic class. The spectrum of 24941 (1997 JM14) is a limiting case for a V-type asteroid, if the relative reflectance at 0.9 $\mu$m were any larger then it would be classified as an S-type. The one candidate that turned out to be an S-type was 46690 (1997 AN23). Even with the low S/N of its spectrum it is clear that the 1 $\mu$m absorption feature is not as deep as that of a typical V-type. As discussed in \S\ref{sec.disc} there is some uncertainty about whether 10537 (1991 RY16) is a basaltic asteroid, but we consider the observation of this target a successful confirmation of a V-type because of the depth of its 1 $\mu$m band.

\section{Size and Orbital Distributions of Basaltic Asteroids \label{sec.dist}}

Using the results and construct of our survey we develop a simple methodology to determine unbiased number distributions of basaltic asteroids that are not part of the dynamical Vestoid population. We calculate a differential size-frequency distribution, $N(H)$ where $H$ is absolute magnitude (a proxy for size), and a semi-major axis distribution, $N(a; H)$. We use $N(H)$ to place estimates on the mass of basaltic material represented by our survey. These calculations are performed by correcting our candidate distributions for biases inherent to the selection process and the SDSS catalog. The underlying assumption for these calculations is that the 11 basaltic candidates that have been observed are taxonomically representative of the other 39.

\subsection{Size-Frequency Distribution}

We begin by equating the expected number of basaltic asteroids in our survey to the number represented by the observations:
\begin{equation}
N(H)~ P_s ~ C(H) = (1-F) ~ n(H),
\label{eqn.NH}
\end{equation}
where $N(H)$ is the unbiased, differential number distribution as a function of absolute magnitude for basaltic asteroids outside of the dynamical Vestoid population. This is the quantity for which we want to solve. $P_s$ is the probability that an asteroid with basaltic surface material would be selected as a candidate based on its photometric properties. $C(H)$ accounts for the completeness and efficiency of the SDSS MOC. Completeness and efficiency are two separate properties. Completeness is the fraction of all asteroids detected by a given survey whereas efficiency is the fraction of observationally accessible asteroids that were detected. The MOC is not 100\% complete for any value of $H$ because it did not have full sky coverage. However, for objects brighter than $H=14.5$ it is 100\% efficient, because it detected all of the objects that were accessible to the survey up to that limiting magnitude. $F$ is the false-positive rate of our survey. Since ten of eleven basaltic candidates have been spectroscopically classified as V-type asteroids (Table \ref{tab.obs}) the false positive rate of basaltic detection, $F=0.09$. As stated in \S\ref{sec.selection} this value may be elevated due to uncertainties in defining the boundaries of the dynamical Vestoid population. For instance, the inner Main Belt ($a<2.5$ AU) candidates may be characterized by one value of $F$ that reflects the scattering of Vestoids to large values of $\Delta v$ while another value characterizes the candidates beyond 2.5 AU. Without additional observations of middle and outer belt candidates we adopt a single value of $F$ for the entire Main Belt and therefore produce an upper limit to the basaltic material represented by the survey. The final parameter in Eq. \ref{eqn.NH}, $n(H)$, is the differential number distribution of our 50 basaltic candidates.

$P_s$ is calculated as the fraction of the 1051 dynamical Vestoids in the MOC that would be selected as candidates based on their photometric colors (this inherently assumes that these dynamical Vestoids are composed of basaltic material). We find that $P_s=0.085$, a low value that is probably related to our stringent constraints across 6 different colors and the relatively large errors on the SDSS photometry.

$C(H)$ is derived by comparing the SDSS MOC to the ASTORB database of all known asteroids. The ratio of the differential absolute magnitude distributions of these two databases gives the product of the efficiency and completeness of the MOC relative to ASTORB. However ASTORB is not 100\% complete past $H\sim14.8$. We extrapolate the number of asteroids for $H>14.8$ with a function of the form $\kappa 10^{\alpha H}$ where $\kappa$ and $\alpha$ are constants \citep{Jedicke02}. Fitting the ASTORB distribution between $H=12.9$ and $H=14.8$ gives the unbiased number of asteroids as a function of absolute magnitude: 
\begin{equation}
N_{all}(H) = (5.9\times10^{-4}) \cdot 10^{0.5 H}.
\label{eqn.fit}
\end{equation}
We limit the extrapolation of this function to a narrow range beyond $H=14.8$. The single slope approximation will slightly overestimate the population of basaltic asteroids because the slope of the distribution is known to decrease at larger $H$ \citep{Jedicke02}.

\begin{figure}[p!]
\begin{center}
\includegraphics[width=8.5cm]{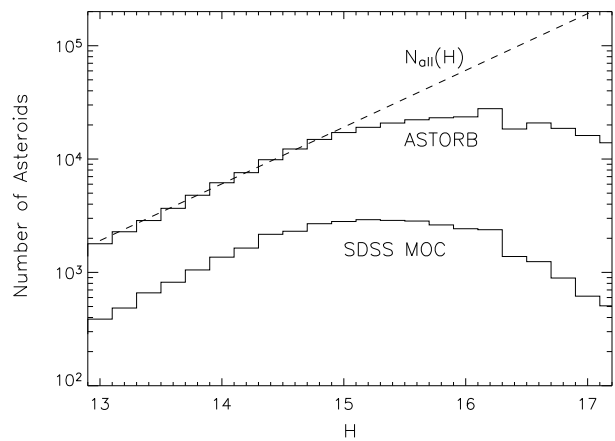}
\caption[]{
Differential number distributions as a function of absolute magnitude for the SDSS MOC and the ASTORB database. The x-axis has been limited to the range of absolute magnitudes represented by our basaltic candidates. The extrapolation of Eq. \ref{eqn.fit} is indicated by the dashed line. The ratio of the SDSS MOC distribution to $N_{all}(H)$ gives the completeness and efficiency of the SDSS as a function of H.}
\label{fig.zoom}
\label{lastfig}
\end{center}
\end{figure}

Our 50 basaltic candidates span a range of absolute magnitudes from 12.9 to 17.2. We focus all of our calculations on the number of basaltic asteroids within this range. Figure \ref{fig.zoom} plots the number distributions of the MOC and ASTORB between $H=12.9$ and 17.2. The fit to the ASTORB data (Eq. \ref{eqn.fit}) is indicated by the dashed line. The ratio of the MOC distribution to $N_{all}(H)$ yields $C(H)$. We find that the combined efficiency and completeness of the MOC is approximately 22\% out to $H=14.5$ and then drops off sharply to less than 1\% at $H=16.5$.

\begin{figure}[p!]
\begin{center}
\includegraphics[width=8.5cm]{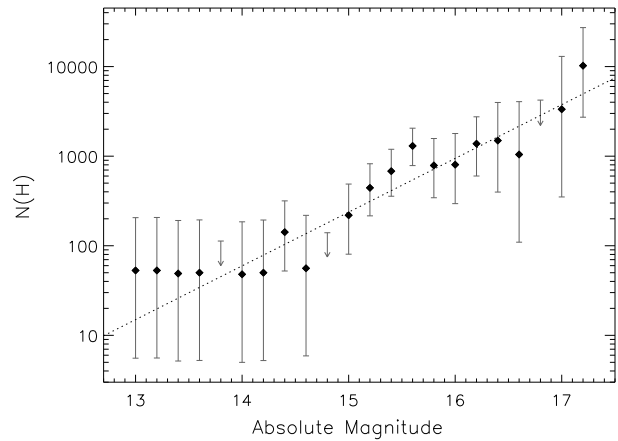}
\caption[]{
Unbiased differential number distribution as a function of absolute magnitude for basaltic asteroids in the Main Belt, excluding those within the dynamical Vestoid population. 90\% Poissonian confidence levels are plotted as upper and lower limits for each point. Certain $H$-bins are empty because they contained no candidates. Upper limits are indicated by arrows for these bins. This distribution is well represented by a single function of the form $\kappa 10^{\alpha x}$ (dotted line).}
\label{fig.NH}
\label{lastfig}
\end{center}
\end{figure}

We solve Eq. \ref{eqn.NH} for $N(H)$ and place upper and lower 90\% confidence levels on the calculated values (Fig. \ref{fig.NH}). The upper limits are equal to the means of the Poisson distributions for which there is a 90\% chance of observing $n(H)+1$ candidates within a given bin, the lower limits are equal to the means for which there is a 90\% chance of $n(H)-1$ candidates. Only upper limits can be calculated for the $H$-bins with no candidates. The apparent flattening of $N(H)$ between $H=13$ and 15 is likely a consequence of low number statistics. 

A single function of the form $\kappa 10^{\alpha x}$ was fit to the $N(H)$ distribution by chi-squared minimization (Fig. \ref{fig.NH}). Extrapolation of this fit suggests that there are no basaltic asteroids  with $H<11$ outside of the dynamical Vestoid population; a result that is confirmed (with the exception of Magnya) by spectroscopic surveys which are close to 100\% complete at $H=10$ and drop to a completeness of $\sim30\%$ at $H=11$ \citep[e.g.][]{Bus02,Lazzaro04}. Combining our calculated $N(H)$ distribution with the findings of these spectroscopic surveys suggests that the overall distribution of basaltic asteroids outside of the dynamical Vestoid population is heavily weighted towards small bodies $< 5$ km in size ($H>13$). The implications of this result are discussed in \S\ref{sec.disc}.

\begin{figure}[p!]
\begin{center}
\includegraphics[width=8.5cm]{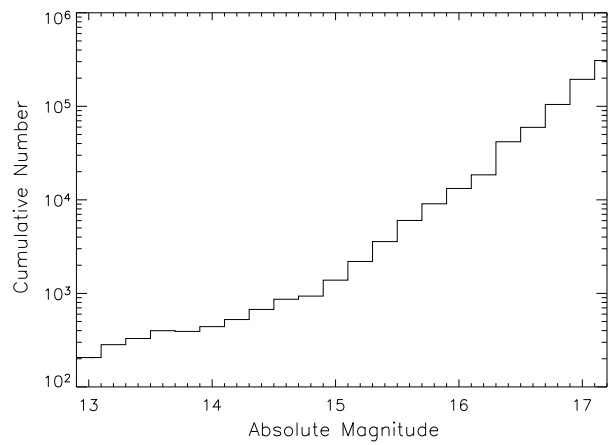}
\caption[]{
Unbiased cumulative number distribution as a function of absolute magnitude for basaltic asteroids in the Main Belt. The value for each bin represents an upper limit to the number of asteroids within the respective absolute magnitude range, thus producing a liberal approximation to the cumulative number.}
\label{fig.Hcum}
\label{lastfig}
\end{center}
\end{figure}

$N(H)$ is used to calculate a cumulative number distribution of basaltic asteroids (Fig. \ref{fig.Hcum}). To place an upper limit on the amount of basaltic material represented by our survey we set the cumulative number of asteroids within a given absolute magnitude range equal to the 90\% Poissonian upper limit. In \S\ref{sec.Na} the unbiased cumulative number of basaltic asteroids is used to normalize the semi-major axis distributions.

\subsection{Masses of Basaltic Material \label{sec.mass}}

We estimate an upper limit to the mass of basaltic material represented in our survey by using the $N(H)$ result. The diameter of an asteroid, $D$, with an assumed geometric albedo, $p$, is \citep{Bowell89}:
\begin{equation}
D = \frac{1347~km}{\sqrt{p}} 10^{-0.2H}.
\label{eqn.D}
\end{equation}
Assuming constant density, $\rho$, and a spherical body we write Eq. \ref{eqn.D} in terms of mass:
\begin{equation}
M(H) = (1.28\times10^{18}~kg) \frac{\rho}{p^{3/2}} 10^{-0.6H}.
\label{eqn.M}
\end{equation}
Thus the total mass represented by our survey is given by:
 \begin{equation}
M_{sur} = \displaystyle \int_{12.9}^{17.2} M(H)~N(H)~dH =  (1.5\times10^{22}~kg) \displaystyle \int_{12.9}^{17.2} N(H)~10^{-0.6H}~dH,
\label{eqn.Mtot}
\end{equation}
where the integral is taken over the range of $H$ represented by our candidates, $\rho$ is set to 3000 kg/m$^3$ and $p$ is assumed to be 0.4, a value that is approximately Vesta's albedo. We solve this integral by integrating over the $N(H)$ upper limits and obtain $M_{sur} = 5.9 \times 10^{16}$ kg. The use of the $N(H)$ upper limits produces a liberal approximation to the total mass represented by our survey. We emphasize that this mass represents the amount of basaltic material within a limited range of absolute magnitude. Based on the results of spectroscopic surveys \citep[e.g.][]{Bus02,Lazzaro04} we expect there to be few undiscovered basaltic asteroids with $H<12.9$, thus the lower H-limit to our calculation is not expected to significantly impact the mass result. Beyond the upper H-limit there may be enough basaltic material to increase $M_{sur}$ by a factor of a few depending on the number distribution at large H.

To place $M_{sur}$ in perspective, we conservatively estimate the mass of basaltic material excavated from the surface of Vesta (based on the size of its largest crater) and place an upper limit on the mass of basaltic material represented by the dynamical Vestoids ($M_V$). As we show in \S\ref{sec.disc} this upper limit to the total mass of basaltic material in the Main Belt ($M_{sur}$ + $M_{V}$) is still much less than that expected from the volume of Vesta's largest crater.

The volume of Vesta's largest crater is estimated to be $10^6$ km$^3$ \citep{Thomas97} and Vesta's mean density is 3.3 g/cm$^3$ \citep{Viateau01}. This volume of material was excavated anywhere from 1 - 3.5 Gyr ago \citep{Bottke05a,Marzari96} and since then has been depleted by collisional grinding and the scattering of fragments out of the asteroid belt. Such depletion involves a complicated interplay of processes whose efficacy is dependent on the size distribution of ejected fragments. Detailed collisional and dynamical models are necessary to address this process. \citet{Nesvorny08} show that $\sim10\%$ of Vestoids would dynamically escape the asteroid belt over a time scale of 2 Gyr. Based on the results of \citet{Bottke05b} we estimate that 50\% is a reasonable estimate for the collisional depletion of the observable Vestoid family (sizes $>1$ km) over a lifetime of 3.5 Gyr. Thus we expect the combined collisional and dynamical depletion of the excavated crater material to be approximately 60\%. Accounting for this depletion, we calculate the remaining mass of excavated basaltic material, $M_{ex}= 1.3\times10^{18}$ kg. 

It is unlikely that a significant fraction of $M_{ex}$ is non-basaltic. The excavation depth, based on the size of Vesta's largest crater, was approximately 13 km \citep{Thomas97}, which is less than or comparable to calculations for the thickness of Vesta's basaltic crust \citep{Keil02} and is comparable to the size of the largest basaltic Vestoids, 1929 Kollaa \citep{Kelley03} and 2045 Peking \citep{Bus02}.

\begin{figure}[p!]
\begin{center}
\includegraphics[width=8.5cm]{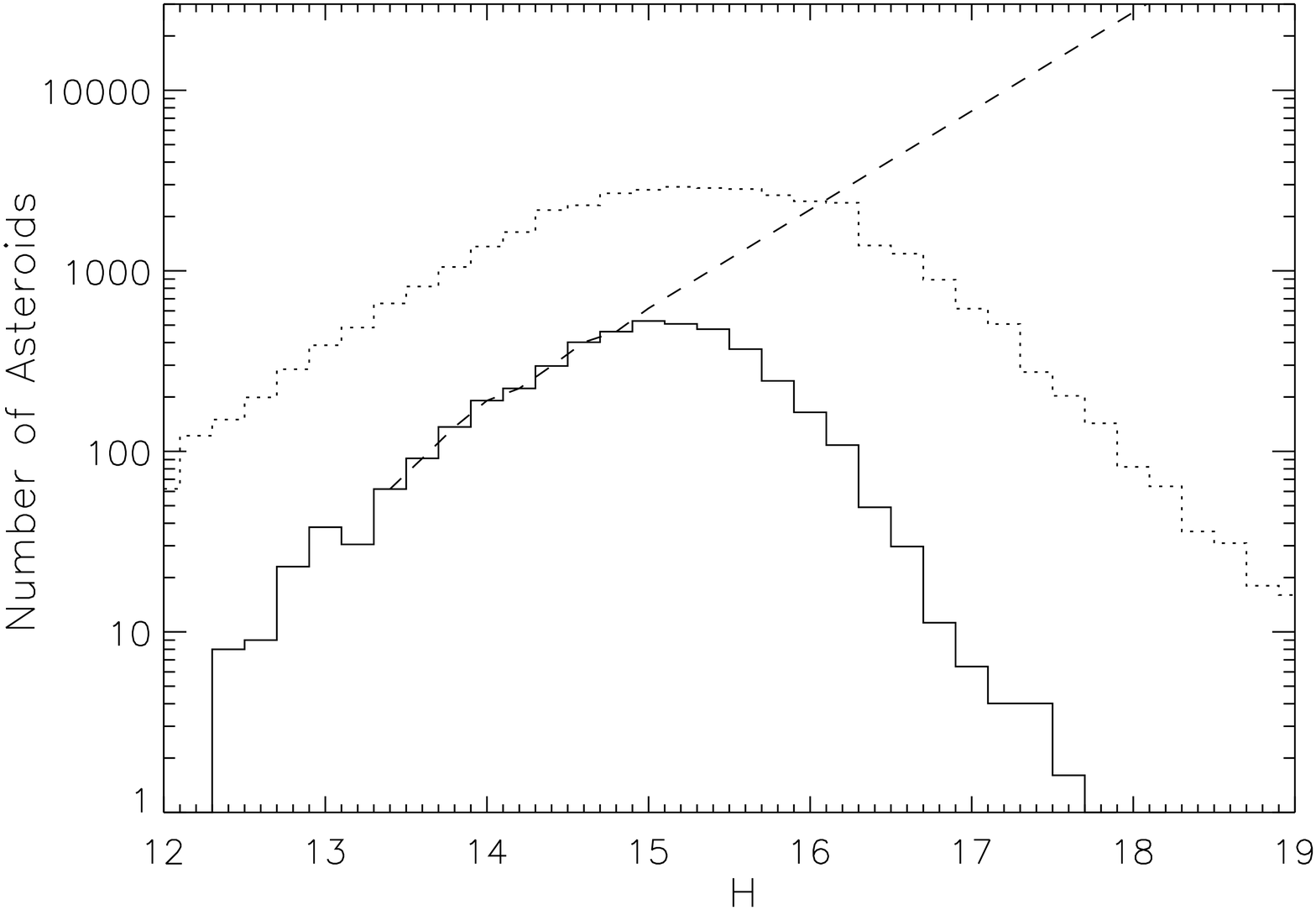}
\caption[]{
Number distribution of dynamically defined Vestoids as a function of absolute magnitude (solid). The fit to the distribution for $13.4 < H < 14.8$ is indicated by the dashed line. The dotted curve is the SDSS MOC distribution. Note the increased deficiency of small bodies ($H > 16$) in the Vestoid family relative to the MOC.}
\label{fig.NvH}
\label{lastfig}
\end{center}
\end{figure}

We calculate the net mass of basaltic material in our list of 5575 dynamical Vestoids by de-biasing this population and removing non-V-type interlopers. Table \ref{tab.interlopers} lists interlopers in the Vestoid dynamical space that are spectroscopically confirmed to be non-V-type and have $H<13$. Objects fainter than this typically do not have an assigned taxonomy, thus we assume that the fraction of interlopers for $H>13$ is $\sim20\%$, the same as the percentage of interlopers for $H<13$. The absolute magnitude distribution for dynamical Vestoids with non-V-type interlopers removed is shown in Fig. \ref{fig.NvH}.

Our list of dynamical Vestoids is derived from the AstDys database which is complete to an absolute magnitude of $H=14.8$. We extrapolate the number of basaltic dynamical Vestoids for $H>14.8$ using the same methodology as in \S5.1. The fit to the dynamical Vestoid distribution from $13.4<H<14.8$ is shown as the dashed line in Fig. \ref{fig.NvH}. The total basaltic mass in the dynamical Vestoid population, $M_V$, is calculated with Eq. \ref{eqn.Mtot} by replacing $N(H)$ with the differential number distribution of dynamical Vestoids, $N_V(H)$, which is a combination of the interloper-corrected number of objects for $N_V(H<14.8)$ and the fitted number of objects for $N_V(H>14.8)$. The limits of integration are changed to 11.9 and 18.0 so that all known dynamical Vestoids are included. The result is $M_{V} = 4.8 \times 10^{16}$ kg.

A few issues may have slightly affected the calculation of $M_{V}$. First, a single-slope fit to the dynamical Vestoid distribution (dashed line in Fig. \ref{fig.NvH}) likely produced an overestimate because unbiased absolute magnitude distributions are expected to turnover at large $H$ \citep{Jedicke02}. Second, the use of a more recent version of the AstDys database would produce a dynamical Vestoid family with more members \citep{Nesvorny08,Roig08} and thus could slightly increase the completeness limit of this population. However, the additional dynamical Vestoids in an updated database would have $H>14.8$ and therefore it is unlikely that they would greatly influence the mass estimate which is based on a reasonable extrapolation for the number of objects at values of $H>14.8$. Lastly, $M_V$ may be slightly overestimated due to unidentified interlopers in the dynamical Vestoid orbital space. All of these issues are expected to produce minor effects and as a whole are expected to increase the upper limit on $M_V$. We discuss the calculated masses, $M_{sur}$, $M_{ex}$ and $M_{V}$, in \S\ref{sec.disc}.

\subsection{Semi-major Axis Distribution \label{sec.Na}}

To calculate the semi-major axis distribution of basaltic asteroids we apply the same methodology as that used to derive the absolute magnitude distribution, i.e.~we simply rewrite Eq. \ref{eqn.NH} in terms of the independent variable $a$, the semi-major axis and $H$:
\begin{equation}
N(a;H)  = A(H) ~\frac{(1-F) ~ n(a)}{P_s ~ C(a,H)},
\label{eqn.Na}
\end{equation}
where $A(H)$ is a normalization factor determined from the $N(H)$ cumulative number distribution. $A(H)$ is set so that the integrated number of basaltic asteroids across all semi-major axes is in agreement with the absolute magnitude result.

The completeness and efficiency of the MOC as a function of semi-major axis, $C(a,H)$, is also dependent on absolute magnitude. For all of the basaltic asteroids represented by our survey:
\begin{equation}
C(a;~H<17.2)  = \frac{n_{SDSS}(a;~H<17.2)}{N_{AST}(a;~H<17.2)},
\label{eqn.Ca}
\end{equation}
where $n_{SDSS}$ is the observed semi-major axis distribution of asteroids in the SDSS MOC with $H$ less than the specified limit and $N_{AST}$ is the unbiased semi-major axis distribution from the ASTORB catalog. However, the latter is only unbiased for $H<14.8$. We calculate $N_{AST}(a;~H<17.2)$ by assuming it is a scaled version of  the $N_{AST}(a;~H<14.8)$ distribution. The scaling is performed so that the integrated number of asteroids represented by $N_{AST}(a;~H<17.2)$ is equal to the cumulative number predicted by $N_{all}$ (Eq. \ref{eqn.fit}). An identical procedure is followed to calculate $C(a,H)$ for other absolute magnitude limits.

\begin{figure}[p!]
\begin{center}
\includegraphics[width=8.5cm]{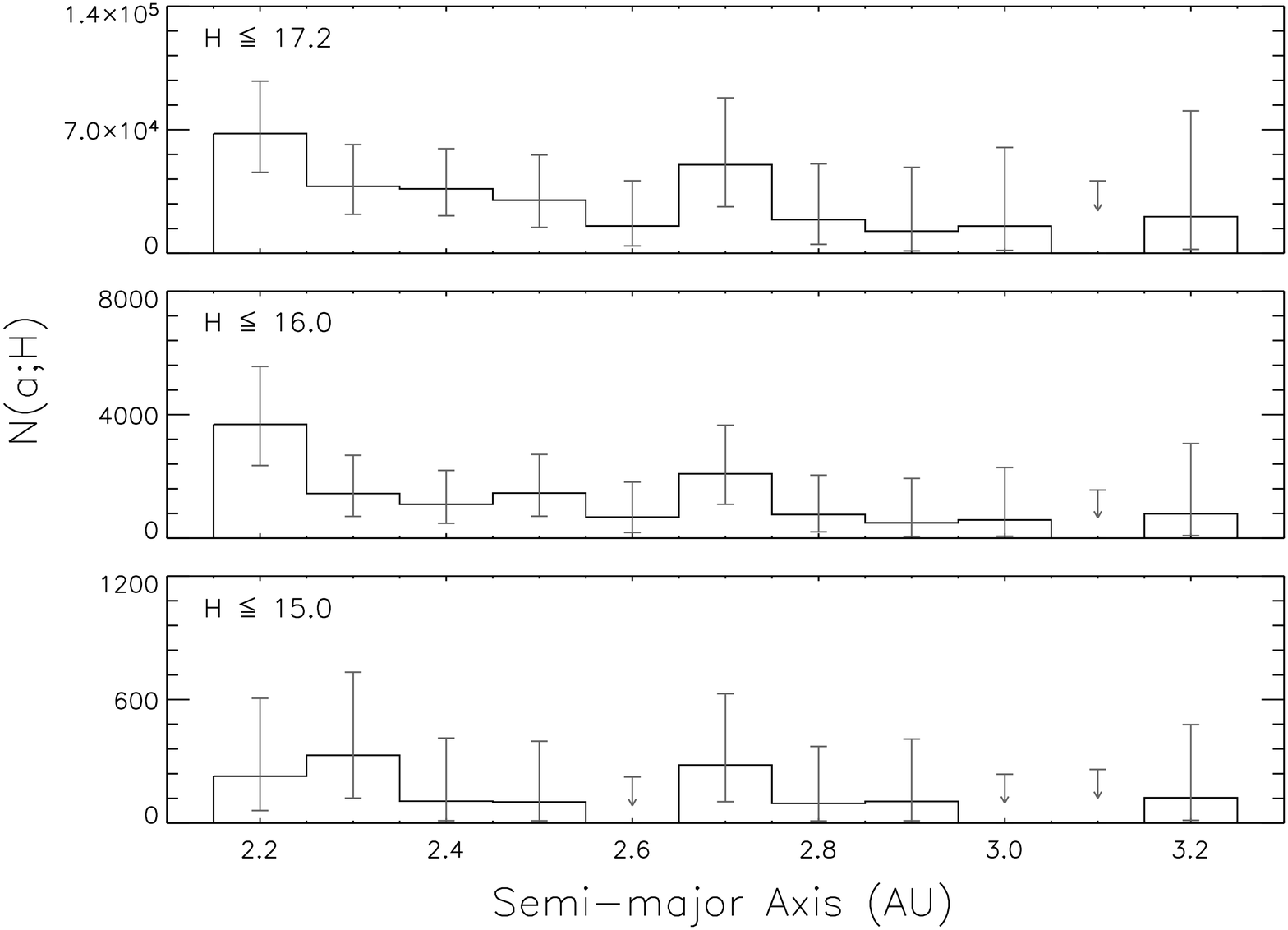}
\caption[]{
Number of basaltic asteroids as a function of semi-major axis and absolute magnitude. The integrated number of objects has been scaled to equal those represented by Fig. \ref{fig.Hcum}. Three different absolute magnitude regimes are shown. The upper and lower limits are calculated from 90\% Poisson confidence levels. For bins without any candidates $N(a;H)$ becomes zero. Upper limits are indicated with arrows for these bins. Assuming a typical Vestian albedo of 0.4, the three absolute magnitude limits correspond to sizes of 0.76 km, 1.3 km and 2.0 km (from top to bottom). Dynamical simulations \citep{Roig08,Nesvorny08} predict that Vestoids smaller than 3 km can cross the 3:1 resonance at 2.5 AU.}
\label{fig.Na}
\label{lastfig}
\end{center}
\end{figure}

Figure \ref{fig.Na} shows the unbiased semi-major axis distributions, $N(a;H)$, for three different absolute magnitude limits. 90\% Poissonian confidence levels are plotted as the upper and lower limits. Some of the semi-major axis bins have zero-value due to the scarcity of candidates with $H<15.0$ and $a>2.5$ AU. The apparent overlap of these non-Vestoid basaltic asteroids with the region occupied by the dynamical Vestoids is a consequence of projection in inclination and eccentricity spaces. 

Figure \ref{fig.Na} allows direct comparison to the results of \citet{Roig08}. These authors calculate size-frequency distributions for the possible number of Vestoids that have migrated across the 3:1 resonance into dynamically stable orbits (see their Fig. 6). In their most optimistic simulations they find that on the order of $100$ Vestoids with a size of 2 km or greater could cross this resonance over the maximum lifetime of the Vesta family (3.5 Gyr). Our lower and upper limits suggest that between 140 and 2800 basaltic asteroids with a size bigger than 2 km ($H < 15.0$) exist beyond 2.5 AU. The lower end of this estimate is commensurate with the \citet{Roig08} result. However, if the basaltic asteroids in the outer belt are exclusively of Vestian origin then they would be distributed at semi-major axes near the 3:1 resonance. Our calculated distribution is in contrast to this expectation: it is roughly uniform across all $a>2.5$ AU and has a single peak at 2.7 AU. This peak might be related to basaltic asteroids generated from a partially differentiated Eunomia parent body \citep[see \S\ref{sec.disc}.1,][]{Reed97,Nathues05}. Ignoring this peak, the flat ``background" population beyond 2.5 AU is suggestive of multiple sources of basaltic asteroids in addition to those few added by the diffusion of Vestoids across the 3:1 resonance. The very low number statistics in this region make it difficult to make more detailed inferences. Further observations are needed to provide more robust constraints on the distribution of these objects beyond the 3:1 resonance.

\section{Discussion and Summary \label{sec.disc}}

We have identified 50 candidate basaltic asteroids from the SDSS MOC based on photometric and dynamical constraints. Based on observations of these basaltic candidates we have constrained the amount of basaltic material that is present outside of a region occupied by the dynamical Vestoids (defined by $a$-$e$-$i$ orbital elements, \S\ref{sec.selection}). We have determined the unbiased distribution of this basaltic material in absolute magnitude and semi-major axis.

\subsection{Basaltic Asteroids with $a>2.5$ AU}

The prospect of multiple basaltic asteroids with semi-major axes greater than 2.5 AU is particularly compelling due to the unlikelihood of numerous such objects originating from the surface of Vesta and then crossing the 3:1 mean motion resonance \citep{Nesvorny08,Roig08}.

Our selection process identified 12 basaltic candidates between the 3:1 and 5:2 mean motion resonances at 2.5 and 2.8 AU respectively. Six of these are clustered around the Eunomia family (Fig. \ref{fig.orbits}) which is one of the largest families in the Main Belt \citep{Zappala95} and may have derived from a partially or fully differentiated parent body. This claim is supported by a compositional gradient across the surface of Eunomia \citep{Reed97,Nathues05} and a spread of surface compositions amongst members of the Eunomia family \citep{Lazzaro99}, both of which are consistent with a differentiated mineralogy \citep{Gaffey02}. Presumably the basaltic crust of the Eunomia parent body was removed in the large family forming collision and/or by subsequent collisions. Asteroid 21238 (1995 WV7) is a spectroscopically confirmed V-type that may represent some of this lost basaltic crust \citep{Hammergren06,Binzel07,Carruba07}. One of the six candidates in the vicinity of Eunomia, asteroid 40521, is a spectroscopically confirmed V-type, however there is some uncertainty about whether this object originated from the Vesta or Eunomia parent bodies \citep{Carruba07,Roig08}. Detailed numerical simulations will be necessary to investigate whether it is dynamically feasible for the remaining five candidates to have originated from Eunomia. 

It is very unlikely that the selection of 6 candidates in the Eunomia family represents a statistical fluke related to the high density of objects in this region. Approximately 11\% of MOC objects beyond the 3:1 resonance are part of the Eunomia family. Thus, assuming a binomial probability distribution for which there is an 11\% chance of a single candidate being selected at random from the Eunomia family, there is only a 0.7\% probability that 6 of the 16 candidates from beyond the 3:1 resonance would belong to the Eunomia family. 

We have identified only 4 basaltic candidates with semi-major axes greater than 2.8 AU (Fig. \ref{fig.orbits}, Table \ref{tab.candidates}). One of these, asteroid 10537 (1991 RY16), may be classified as a V-type because of its deep 1 $\mu m$ absorption feature which is typical of this class and indicative of a pyroxene-rich and/or olivine-rich mineralogy. However it also has an unusually deep absorption feature centered at $0.63~\mu m$ that is uncommon to V-types (Fig. \ref{fig.RY16}). These visible wavelength data have been confirmed by other observations [Duffard \& Roig (2008), submitted to Planetary and Space Science]. Follow-up infrared observations provide additional insight on the surface composition of this unusual object \citep{Moskovitz08}. Another outer belt candidate is unnumbered asteroid 2003 SG55. It is possible that this object originated from a differentiated Eos parent body \citep{Mothe05}. The presence of multiple non-linear secular resonances that pass through the region \citep{Milani90,Milani92} may have provided a pathway for the orbital migration of 2003 SG55 even though HCM analysis \citep{Zappala90} shows that 2003 SG55 is outside of the Eos dynamical family. Future study of these outer belt basaltic candidates will have important implications for understanding the dynamical and thermal histories of these specific families.

\subsection{Basaltic Asteroids with $a<2.5$ AU}

Defining the dynamical boundaries of the Vestoids is difficult due to effects such as Yarkovsky drift and resonance scattering \citep{Carruba05,Carruba07}. Nevertheless we have attempted to identify non-Vestoid basaltic candidates inside of 2.5 AU because neglecting to do so could exclude basaltic asteroids of non-Vestian origin. The relatively uniform distribution of basaltic candidates in this region suggests two possibilities: (1) there exists non-Vestoid basaltic asteroids in the inner belt that represent a unique differentiated parent body, or (2) Vestian fragments actually populate the full dynamical extent of the inner Main Belt. The first possibility is testable by performing mineralogical analyses (e.g.~\citet{Lawrence07,Gaffey93a}) on NIR spectra of basaltic candidates with large values of $\Delta v$, for example 38070 (1999 GG2) with $\Delta v=2.1$ km/s and 33881 (2000 JK66) with $\Delta v=3.3$ km/s. If these inner belt basaltic candidates do represent a unique differentiated parent body then they may be compositionally distinguishable from Vesta and the Vestoids (e.g.~Magnya, \citet{Hardersen04}). The second possibility can be investigated with numerical simulations. \citet{Nesvorny08} show that the dynamical diffusion of Vestian material does in fact populate the semi-major axis extent of the inner Main Belt, but does not extend below orbital inclinations of 4$^\circ$. In the future, the most effective test for inner belt, non-Vestoid basaltic asteroids will be a combination of dynamical studies and mineralogical analyses.

\subsection{Total Basaltic Asteroid Inventory}

We find that the calculated size-frequency distribution of basaltic asteroids outside of the dynamical Vestoid population is heavily weighted towards small bodies. We have also shown that the mass of basaltic material outside of the dynamical Vestoid population is $M_{sur} < 5.9 \times 10^{16}$ kg. This value is comparable to the mass of basaltic material inside of the dynamical Vestoid population. Due to uncertainties in the definition of dynamical families we do not attempt to make a statement about the fraction of $M_{sur}$ that is truly non-Vestoid in origin. We have attempted to place upper limits on $M_{sur}$ and $M_V$ and a conservative estimate on $M_{ex}$. Even so we note that $M_{sur} + M_V < M_{ex}$ by over an order of magnitude.

The size-frequency distribution ($N(H)$) and smaller total mass relative to $M_{ex}$, both inclusive and exclusive to the Vestoid dynamical family, are consistent with most (but not all!) of the basaltic asteroids deriving from the Vesta parent body. Vesta has several large craters on its surface \citep{Thomas97} whose collisional formation would have produced fragments of varying size. Due to ejection velocities, the Yarkovsky effect, resonance interactions and close encounters, these fragments diffused away from the Vesta parent body \citep{Bottke01,Carruba05,Nesvorny08}. The smallest fragments experienced the fastest dynamical evolution while larger objects would take longer to separate from Vesta due to the decreased significance of the Yarkovsky force for bodies larger than 10 km \citep{Bottke00}. Over time, small Vestoids migrated to the edge of the family at the $\nu_6$ and 3:1 resonances. Although many of the fragments that reached these resonances were scattered into planet crossing orbits or into the Sun, some crossed over into stable orbits \citep{Gladman97,Roig08,Nesvorny08}. Thus, the size-frequency distribution of Vestoids shows a depletion of small bodies ($H>16$) relative to the SDSS MOC distribution. This is apparent in Fig. \ref{fig.NvH} where  the slope of the Vestoid distribution (solid curve) for $H>16$ is slightly steeper than that of the MOC (dotted curve). This also holds true when compared to just the subset of MOC objects that are confined to the semi-major axis range of the Vestoids ($2.2 < a < 2.5$ AU). 

The overabundance of small bodies (sizes $< 2$ km) amongst the basaltic asteroids outside of the dynamical Vestoid population (Fig. \ref{fig.NH}) may represent the hiding place of these "missing Vestoids". If this is true then it suggests that the shallow slope of absolute magnitude distributions at large values of $H$ for HCM-defined collisional families \citep{Morby03} may be attributed to the preferential loss of small bodies out of the families by dynamical processes such as the Yarkovsky effect and resonance scattering.

The semi-major axis distribution indicates that the full inventory of basaltic material in the Main Belt is not consistent with an origin exclusively coupled to the formation of the large Vestian crater. With an age of 1-3.5 Gyr \citep{Bottke05a,Marzari96}, the primary Vesta crater was formed after the completion of planet formation. Thus, the asteroidal fragments generated in this impact evolved in a dynamical environment similar to the present day architecture of secular and mean motion resonances. In such an environment it is relatively easy for asteroids to migrate in semi-major axis because of the Yarkovsky effect, however it is more difficult to produce significant changes in inclination and eccentricity. This motivates the search for basaltic asteroids at very different $i$ and $e$ values than Vesta. 

Our results suggest that basaltic asteroids with sizes in excess of 2 km do exist at low inclinations and beyond 2.5 AU (Fig. \ref{fig.Na}). \citet{Roig08} show that it is not possible to dynamically transport many hundreds of Vestoids with diameters greater than 2 km beyond the 3:1 resonance. This is further confirmed by \citet{Nesvorny08} whose dynamical simulations do not produce any 3 km size bodies beyond this resonance or any Vestoids with low inclination orbits ($i<4^\circ$). 

We cite three explanations for the presence of basaltic material in regions of phase space that are dynamically inaccessible to the Vestoids. One possibility is that these basaltic asteroids originated from the surface of Vesta, not from its largest crater but sometime before the Late Heavy Bombardment (LHB) more that 3.8 Gyr ago. The higher density of objects in the pre-LHB asteroid belt \citep{Petit02} would have made for a dynamical environment in which collisional and scattering events were greatly enhanced over present day rates. In addition, resonance sweeping during the LHB could have dispersed these asteroids' inclinations \citep{Nesvorny08}.

A second possibility is that these objects represent scattered planetesimals from the terrestrial planet region of the Solar System \citep{Bottke06}. In the inner Solar System solid-body accretion times would have been faster than in the Main Belt. These bodies would have accreted a higher quantity of live radioactive nuclides such as $^{26}$Al and been more likely to differentiate. The amount of basaltic material that was injected into the Main Belt from the inner Solar System is currently unconstrained. 

A final explanation for the presence of basaltic asteroids at low inclinations and beyond 2.5 AU is that they derived from other differentiated parent bodies that formed in the Main Belt. Although we find no asteroidal evidence for a large number of undiscovered basaltic asteroids, we cannot rule out minor contributions to the basaltic inventory from non-Vestoid differentiated bodies whose basalt has been heavily depleted due to collisions and dynamical effects \citep[i.e. a "battered to bits" scenario,][]{Burbine96}. It is possible that the observed distribution of basaltic asteroids in the Main Belt is derived from many different source populations including the Vestoids, inner Solar System planetesimals and a heavily depleted population of non-Vestoid differentiated bodies.

\subsection{Comparison to Other Works}

\begin{figure}[p!]
\begin{center}
\includegraphics[width=8.5cm]{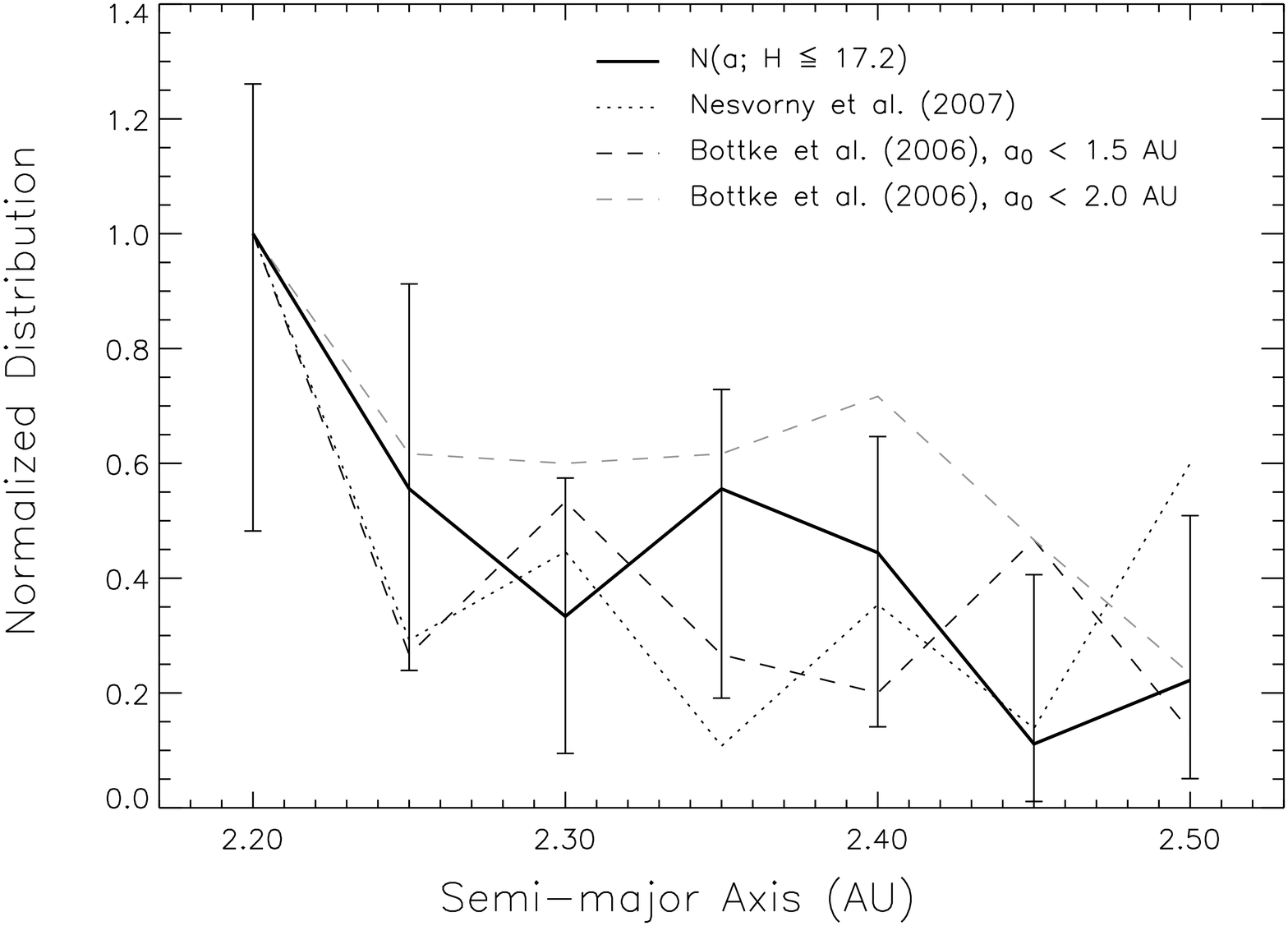}
\caption[]{
Comparison of semi-major axis distributions. Our calculated distribution of basaltic asteroids is shown as the solid line with 90\% Poissonian upper and lower limits. The results of two dynamical simulations are over-plotted. The dotted line shows the simulated number of Vestian fragments that have evolved outside of the HCM-defined dynamical limits of this family \citep{Nesvorny08}. The dashed lines plot the distribution of inner Solar System planetesimals scattered into the Main Belt from two different source regions, $a<1.5$ AU (black dashed line) and $a<2.0$ AU (grey dashed line) \citep{Bottke06}.}
\label{fig.Na_comp}
\label{lastfig}
\end{center}
\end{figure}

We compare our calculated semi-major axis distribution to the simulation results of \citet{Nesvorny08} and \citet{Bottke06} (Fig. \ref{fig.Na_comp}). This allows us to investigate the feasibility of two different scenarios for delivering basaltic asteroids to orbits outside of the dynamical Vestoid population. Figure \ref{fig.Na_comp} shows the normalized distribution of non-Vestoid basaltic asteroids as a function of semi-major axis. These distributions have been normalized by their maximum values (all at 2.2 AU) because limitations inherent to each work precludes the comparison of an absolute number of basaltic asteroids. We confine these distributions to the inner Main Belt ($a<2.5$ AU) because of low number statistics for larger values of $a$ in our analysis and the \citet{Bottke06} simulations and the lack of any Vestoids across the 3:1 resonance in the \citet{Nesvorny08} simulations. The distributions from the numerical simulations exclude test particles that populate the dynamical Vestoid region of orbital element space as defined in \S\ref{sec.selection}. This produces distributions that are non-Vestoid based upon our specific dynamical criteria. 

The \citet{Nesvorny08} study was designed to understand the dynamical evolution of Vesta's ejected fragments. They modeled the diffusion of Vestoids over an estimated family lifetime of 2 Gyr. If our distribution of basaltic asteroids was exclusively derived from dynamically aberrant Vestoids then we would expect it to match the \citet{Nesvorny08} result. 

The \citet{Bottke06} study models the injection of planetesimal-size bodies from the inner Solar System into the Main Belt during the epoch of planet formation. The specific source region of these planetesimals determines which are differentiated and contain basalt. Due to uncertainties in the time-scales of accretion and the initial abundance of radioactive isotopes, it is not clear where the boundary between differentiated and undifferentiated planetesimals existed. Thus we plot two distributions from the \citet{Bottke06} results: one corresponding to all of the planetesimals that are scattered into the main belt from initial semi-major axes, $a_0$, less than 1.5 AU and those with $a_0<2.0$ AU. If our distribution was derived primarily from these scattered interlopers we would expect it to match the \citet{Bottke06} result. 

We calculate the chi-squared ($\chi^2$) probability that these simulated distributions are derived from our observational result. The \citet{Nesvorny08} distribution and the \citet{Bottke06} distribution for a planetesimal source region of $a_0<1.5$ AU produce $\chi^2$ probabilities equal to 99.6\% and 98.5\% respectively. Within the limits of our calculations these distributions fit the data equally well. The worst fit ($\chi^2$ probability = 73\%) is the \citet{Bottke06} distribution for a planetesimal source region of $a_0<2.0$ AU. The difference in probabilities from the \citet{Bottke06} simulations suggests that the contribution of basaltic asteroids from inner Solar System planetesimals is weighted towards objects originally separated in semi-major axis from the Main Belt rather than objects that formed at semi-major axes around 2.0 AU. This is not unexpected as faster accretion times at smaller heliocentric distances would have produced a greater number of differentiated planetesimals.

This work is the first survey specifically designed to constrain the global distribution of basaltic material in the Main Belt. Future all-sky surveys such as PanSTARRS \citep{Kaiser02} and LSST \citep{Claver04} will greatly increase the number of known asteroids and will allow a more complete description of the basaltic asteroids. The results presented here suggest that the total amount of basaltic material in the Main Belt is very much less than expected from the $\sim100$ differentiated parent bodies traced by the iron meteorites and basaltic achondrites, a result which agrees with the suggestion that basaltic fragments have been comminuted to sizes below observational limits \citep[e.g.][]{Burbine96} or that some unknown space weathering processes has obscured the spectral signature of basalt on asteroid surfaces \citep{Wetherill88}. However, the inferred presence of a small amount of basaltic material throughout the Main Belt suggests that some vestige of crustal material from large differentiated bodies does exist and may represent the upper size limit of this population.

\ack
Thanks to Greg Wirth for his patience and expert advice as we have strived for competency with ESI. We would like to acknowledge the tremendous help received from Greg Aldering, Rolin Thomas, Yannick Copin and the rest of the SNIFS team. Thanks to Bobby Bus for helpful suggestions regarding observing and data reduction. We are grateful to Richard Binzel and Fernando Roig for their insightful reviews. N.M. would like to acknowledge the support of NASA GSRP grant NNX06AI30H. R.J. would like to acknowledge the support of NSF planetary astronomy grant AST04-07134. E.G. acknowledges support from the NASA Astrobiology Institute. Some of the data presented herein were obtained at the W.M. Keck Observatory, which is operated as a scientific partnership among the California Institute of Technology, the University of California and the National Aeronautics and Space Administration. The Observatory was made possible by the generous financial support of the W.M. Keck Foundation. Part of the data utilized in this publication were obtained and made available by the The MIT-UH-IRTF Joint Campaign for NEO Reconnaissance. The IRTF is operated by the University of Hawaii under Cooperative Agreement no. NCC 5-538 with the National Aeronautics and Space Administration, Office of Space Science, Planetary Astronomy Program. The MIT component of this work is supported by the National Science Foundation under Grant No. 0506716. We wish to recognize and acknowledge the very significant cultural role and reverence that the summit of Mauna Kea has always had within the indigenous Hawaiian community.  We are most fortunate to have the opportunity to conduct observations from this mountain.

\label{lastpage}


\bibliography{bibliography.bib}

\bibliographystyle{plainnat}


\clearpage	

\begin{table}
\begin{center}
\textbf{}
\begin{tabular}{cc}
\hline 
\hline
Color & Range	\\
\hline
$u-g$ & [0.170,0.350] \\
$g-r$ & [0.201,0.281] \\
$r-i$ & [0.061,0.121]   \\
$i-z$ &  [-0.389,-0.249] \\
PC1 & [0.360,0.520]   \\
PC2 & [-0.258,-0.078] \\
\hline
\end{tabular}
\caption[]{Vestoid Colors -- 50$^{th}$ Percentile}
\label{tab.percentile}
\label{lasttable}
\end{center}
\end{table}

\clearpage

\begin{center}
\begin{longtable}{llccccccc}
\hline 
\hline
Number & Designation & $a$ (AU) & $e$ & sin($i$) & $\Delta v$ (km/s) & RG06? & H06? & B07? \\
\hline

156914 & 2003 FB31 & 2.37 & 0.10 & 0.14 & 0.3 & & & \\
{\bf 28517} & {\bf 2000 DD7} & 2.29 & 0.09 & 0.14 & 0.4 & \checkmark & \checkmark & \checkmark \\
124343 & 2001 QF117 & 2.22 & 0.11 & 0.12 & 0.7 & & \checkmark & \\
112257 & 2002 LC13 & 2.22 & 0.10 & 0.11 & 0.7 & & \checkmark & \\
54668 & 2000 WO85 & 2.44 & 0.11 & 0.10 & 0.7 & & \checkmark & \\
10544 & 1992 DA9 & 2.23 & 0.11 & 0.13 & 0.8 & \checkmark & \checkmark & \\
{\bf 7558} & {\bf Yurlov} & 2.29 & 0.11 & 0.09 & 0.9 & & \checkmark & \\
74894 & 1999 TB119 & 2.22 & 0.12 & 0.14 & 0.9 & & \checkmark & \\
67416 & 2000 QW64 & 2.17 & 0.08 & 0.10 & 1.0 & & \checkmark & \\
65507 & 4151 P-L & 2.22 & 0.12 & 0.11 & 1.0 & \checkmark & \checkmark & \checkmark \\
{\bf 60669} & {\bf 2000 GE4} & 2.21 & 0.13 & 0.12 & 1.0 & \checkmark & \checkmark & \checkmark \\
{\bf 24941} & {\bf 1997 JM14} & 2.48 & 0.12 & 0.09 & 1.0 &  & \checkmark & \\
55804 & 1994 PD13 & 2.29 & 0.13 & 0.15 & 1.1 & & & \\
{\bf 2823} & {\bf van der Laan} & 2.41 & 0.06 & 0.08 & 1.1 & \checkmark & \checkmark & \checkmark \\
{\bf 56570} & {\bf 2000 JA21} & 2.38 & 0.10 & 0.07 & 1.2 & \checkmark & \checkmark & \checkmark \\
98278 & 2000 SE212 & 2.16 & 0.06 & 0.10 & 1.3 & & \checkmark & \checkmark \\
55998 & 1998 SQ135 & 2.19 & 0.10 & 0.08 & 1.3 & \checkmark & \checkmark & \\
{\bf 46690} & {\bf 1997 AN23} & 2.24 & 0.10 & 0.07 & 1.3 & \checkmark & \checkmark & \\
\hspace{0.35cm} - & 2002 RS123 & 2.37 & 0.16 & 0.14 & 1.4 & \checkmark & \checkmark & \checkmark \\
6081 & Cloutis & 2.49 & 0.16 & 0.12 & 1.4 & \checkmark & \checkmark & \\
110544 & 2001 TX96 & 2.68 & 0.06 & 0.17 & 1.6 & & & \\
{\bf 10537} & {\bf 1991 RY16} & 2.85 & 0.10 & 0.11 & 1.6 & \checkmark & \checkmark & \\
122656 & 2000 RV93 & 2.75 & 0.13 & 0.09 & 1.7 & & \checkmark & \\
44114 & 1998 HN21 & 2.18 & 0.02 & 0.10 & 1.7 & & \checkmark  & \\
8805 & 1981 UM11 & 2.34 & 0.12 & 0.05 & 1.7 & \checkmark & \checkmark & \checkmark \\
\hspace{0.35cm} - & 2002 TF149 & 2.77 & 0.02 & 0.10 & 1.9 & & \checkmark & \\
109387 & 2001 QL168 & 2.59 & 0.13 & 0.20 & 2.0 & & \checkmark & \\
86493 & 2000 DD17 & 2.14 & 0.09 & 0.05 & 2.0 & & & \\
158606 & 2002 XS80 & 2.26 & 0.17 & 0.07 & 2.1 & & & \\
{\bf 38070} & {\bf 1999 GG2} & 2.14 & 0.13 & 0.06 & 2.1 & & & \\
\hspace{0.35cm} - & 2001 QA244 & 2.73 & 0.18 & 0.17 & 2.2 & & & \\
93322 & 2000 SA221 & 2.74 & 0.10 & 0.21 & 2.2 & & \checkmark & \\
84036 & 2002 PF50 & 2.33 & 0.15 & 0.04 & 2.2 & \checkmark & \checkmark & \\
55092 & 2001 QO123 & 2.15 & 0.13 & 0.05 & 2.2 & & \checkmark & \\
69634 & 1998 FH68 & 2.38 & 0.16 & 0.05 & 2.3 & & \checkmark & \\
41910 & 2000 WS141 & 2.26 & 0.20 & 0.13 & 2.3 & \checkmark & \checkmark & \\
{\bf 40521} & {\bf 1999 RL95} & 2.53 & 0.05 & 0.22 & 2.3 & \checkmark & \checkmark & \\
152169 & 2003 AV27 & 2.25 & 0.10 & 0.02 & 2.4 & & & \\
107008 & 2000 YM112 & 2.35 & 0.18 & 0.05 & 2.4 & & \checkmark & \\
\hspace{0.35cm} - & 2003 SG55 & 3.01 & 0.13 & 0.20 & 2.5 & & \checkmark & \\
50802 & 2000 FH27 & 2.91 & 0.09 & 0.03 & 2.5 & & \checkmark & \\
140706 & 2001 UV79 & 2.61 & 0.03 & 0.02 & 2.6 & & \checkmark & \\
55550 & 2001 XW70 & 2.55 & 0.14 & 0.23 & 2.6 & & \checkmark & \\
172731 & 2004 BY120 & 2.74 & 0.14 & 0.23 & 2.8 & & & \\
129632 & 1998 HV36 & 2.53 & 0.24 & 0.12 & 3.0 & & & \\
84021 & 2002 PC41 & 2.68 & 0.09 & 0.25 & 3.0 & & \checkmark & \\
{\bf 33881} & {\bf 2000 JK66} & 2.21 & 0.24 & 0.18 & 3.3 & \checkmark & \checkmark & \\
86627 & 2000 EJ126 & 2.42 & 0.24 & 0.04 & 3.5 & & & \\
\hspace{0.35cm} - & 2001 GQ5 & 2.20 & 0.26 & 0.15 & 3.6 & & & \\
111515 & 2001 YC91 & 3.16 & 0.16 & 0.28 & 3.9 & & \checkmark & \\
\hline
\caption[]{Non-Vestoid Basaltic Candidates\\
The table columns are: asteroid number and designation, semi-major axis, eccentricity,  
sine of inclination, ejection velocity ($\Delta v$) required to have originated from Vesta \citep{Zappala96} and the last three columns indicate whether the asteroid has been identified as a V-type candidate by \citet[][RG06]{Roig06}, \citet[][H06]{Hammergren06} and \citet[][B07]{Binzel07}. The candidates are sorted by their $\Delta v$ values. Spectroscopic data have been obtained for candidates listed in bold-face (Table \ref{tab.obs}, Fig. \ref{fig.opt_spec}). The orbital elements of Vesta are: $a = 2.36$ AU, $e = 0.09$ and sin$(i) = 0.12$.
}
\label{tab.candidates}
\end{longtable}
\end{center}

\begin{table}
\begin{center}
\textbf{}
\begin{tabular}{lccccccc}
\hline 
\hline
Object & UT Date & Inst. & Mag. & $t_{exp}$ (s) & Analog & Type\\
\hline
7558 (Yurlov) & Oct. 1, 2006 & ESI & 16.1 & 1800 & Avg. & V\\
10537 (1991 RY16) & Oct. 1, 2006 & ESI & 17.9 & 2700 & Avg. & V?\\
 & Jan. 18, 2007 & SNIFS & 18.3 & 3600 & SA93-101& V?\\
24941 (1997 JM14) & Oct. 1, 2006 & ESI & 19.0 & 900 & Avg.& V\\
28517 (2000 DD7) & Mar. 10, 2007 & ESI & 16.5 & 900 & Avg. & V \\
33881 (2000 JK66) & Nov. 23, 2007 & SpeX & 16.6 & 2880 & HD9562 & V \\
38070 (1999 GG2) & Aug. 23, 2006 & ESI & 17.8 & 2700 & Avg. & V\\
& Oct. 05, 2006 & SpeX & 16.9 & 1920 & HD377 & V\\
46690 (1997 AN23) & Mar. 10, 2007 & ESI & 18.7 & 2700 & Avg. & S \\
56570 (2000 JA21) & Jan. 4, 2006 & ESI & 19.0 & 2700 & Avg. & V\\
60669 (2000 GE4) & Jun. 18, 2007 & SNIFS & 17.8 & 4500 & BS5534 & V\\
2823 (van der Laan)$^1$ & Nov. 22, 2005 & SpeX & 16.4 & 480 & SA93-101 & V\\
40521 (1999 RL95)$^2$ & Apr. 30, 2006 & GMOS & 18.1 & 3000 & SA107-871 & V\\
\hline
\end{tabular}
\caption[]{Observation Summary\\
The columns in this table are: object number and designation, UT date of observation, the instrument used, magnitude of the target from the JPL HORIZONS system, net exposure time in seconds, the solar analog that was used for calibration in the reduction process (see \S\ref{sec.reduct}, Avg. indicates that the average of all solar analogs for a given night were used) and the taxonomic designation that we assigned to the spectra based upon the criteria outlined by \citet{Bus02}. \\
$^1$SMASS Database (http://smass.mit.edu/minus.html), IRTF\\
$^2$\citet{Roig08}, Gemini South
}
\label{tab.obs}
\label{lasttable}
\end{center}
\end{table}

\begin{table}
\begin{center}
\textbf{}
\begin{tabular}{lccc}
\hline 
\hline
Object & H & Type & Reference	\\
\hline
63 (Ausonia) & 7.25 & S & (1) \\
556 (Phyllis) & 9.23 & S & (1)\\
1145 (Robelmonte) & 11.05 & TDS & (2)\\
2098 (Zyskin) & 11.92 & TX & (2)\\
2086 (Newell) & 11.99 & Xc & (1)\\
2346 (Lilio) & 12.04 & C & (1)\\
3376 (Armandhammer) & 12.33 & Sq & (1)\\
3865 (1988 AY4) & 12.42 & Xc & (1)\\
4510 (Shawna) & 12.69 & S & (2)\\
2024 (McLaughlin) & 12.73 & S & (2)\\
1781 (van Biesbroeck) & 12.75 & XS & (2)\\
5111 (Jacliff) & 12.78 & R & (1)\\
4845 (Tsubetsu) & 12.81 & X & (1)\\
1697 (Koskenniemi) & 12.93 & TX & (2)\\
5600 (1991 UY) & 12.94 & S & (3)\\
\hline
\end{tabular}
\caption[]{Interlopers in Vestoid Dynamical Space\\
(1) \citet{Bus02}\\
(2) \citet{Xu95}, uncertain classifications are listed in order of taxonomic likelihood\\
(3) \citet{Lazzaro04}\\
}
\label{tab.interlopers}
\label{lasttable}
\end{center}
\end{table}

\clearpage


\clearpage

\clearpage

\clearpage

\clearpage

\clearpage

\clearpage

\clearpage

\clearpage

\clearpage

\clearpage

\clearpage

\end{document}